\documentclass[preprint,11pt,authoryear]{elsarticle}
\usepackage{iftex}
\ifPDFTeX
  \usepackage[T1]{fontenc}
  \usepackage[utf8]{inputenc}
  \usepackage{tgpagella}
\else
  \usepackage{fontspec}
\fi
\usepackage{newpxtext}
\usepackage{geometry}
\usepackage{amsmath}
\usepackage{amsfonts}
\usepackage{siunitx}
\usepackage{amssymb}
\usepackage{amsthm}
\usepackage{amsopn}
\usepackage{mathtools}
\usepackage{bm}
\usepackage{bbm}
\usepackage[version=4]{mhchem}
\usepackage{graphicx}
\usepackage{subcaption}
\usepackage{rotating}
\usepackage{pdflscape}
\usepackage{svg}
\usepackage{pgfplots}
\pgfplotsset{compat=1.18}

\usepackage{threeparttable}
\usepackage{tabularx}
\usepackage{tabularray}
\usepackage{array}
\usepackage{booktabs}
\usepackage{multirow}
\usepackage{makecell}
\usepackage{colortbl}
\usepackage{arydshln}

\usepackage{float}
\usepackage{enumitem}
\usepackage{framed}
\usepackage{tcolorbox}
\usepackage{listings}

\usepackage{algorithm}
\usepackage{algpseudocode}

\usepackage{xcolor}
\usepackage[normalem]{ulem}
\usepackage[makeroom]{cancel}
\usepackage[symbol]{footmisc}

\usepackage{url}
\usepackage{doi}

\usepackage{lineno}
\usepackage{cleveref}
\usepackage{hyperref}

\setcitestyle{numbers,square}
\usepackage{pifont}
\newcommand{\xmark}{\ding{55}}
\newcommand{\rom}[1]{\expandafter{\romannumeral #1\relax}}

\theoremstyle{definition}

    \newcommand{\opid}[1]{\texttt{OpModeID #1}}
\newcount\Comments  %
\newcommand{\kibitz}[2]{\ifnum\Comments=0\textcolor{#1}{#2}\fi}

\newcommand{\sectionname}{Section}

\usepackage{lineno}
\journal{arXiv.org}
\begin{document}
\begin{frontmatter}
\title{\textbf{\Large Emission reduction potential of freeway stop-and-go wave smoothing}}
\author[inst1,inst2]{Junyi Ji}
\ead{junyi.ji@vanderbilt.edu}
\author[inst2]{Derek Gloudemans}
\author[inst2]{Gergely Zach\'ar}
\author[inst2]{William Barbour}
\author[inst1,inst2,inst3]{\\Jonathan Sprinkle}
\author[inst4]{Benedetto Piccoli}
\author[inst1,inst2,inst3]{Daniel B. Work}
\affiliation[inst1]{organization={Department of Civil and Environmental Engineering, Vanderbilt University},country={United States}}
\affiliation[inst2]{organization={Institute for Software Integrated Systems, Vanderbilt University},country={United States}}
\affiliation[inst3]{organization={Department of Computer Science, Vanderbilt University},country={United States}}
\affiliation[inst4]{organization={Department of Mathematical Sciences, Rutgers University–Camden},country={United States}}
\begin{abstract}
The real-world potential of stop-and-go wave smoothing at scale remains largely unquantified. Smoothing freeway waves requires opening a gap large enough for them to dissipate, but that gap is often impractically large. We propose a counterfactual wave-smoothing benchmark that reconstructs a smooth, feasible trajectory from each empirical trajectory by solving a quadratic program with fixed boundary conditions and a maximum-gap constraint, and we use the MOVES model to estimate the resulting emission reduction potential. Applying the framework to nine weeks of weekday peak-period data from the I-24 MOTION testbed, which exhibits rich day-to-day variation in wave dynamics, we find meaningful potential for passenger cars under a 0.1-mile maximum-gap constraint: average \ce{CO2} reductions of 9.80\% to 14.08\% across lanes, with 12.94\% to 25.70\% in \ce{CO}, 24.29\% to 29.76\% in \ce{HC}, and 29.40\% to 36.03\% in \ce{NOx}. Trucks emerge as particularly attractive targets, showing roughly twice the reduction potential of passenger cars.
\end{abstract}

\begin{keyword}
stop-and-go waves \sep emissions \sep traffic smoothing \sep freeway traffic
\end{keyword}

\end{frontmatter}

\section{Introduction}
Freeway stop-and-go waves~\cite{edie1967generation,kwon2000day,bertini2005congestion,laval2010mechanism,ji2026scalable} are collective, repeated speed oscillations that emerge and propagate upstream, against the direction of traffic flow~\cite{lighthill1955kinematic,helbing1998jams,ahn2005formation}, forcing vehicles into inefficient acceleration cycles and increasing fuel use and pollutant emissions~\cite{barth2008real}. Decades of research on active traffic control~\cite{papageorgiou2003review,ma2016freeway,vishnoi2024cav,lee2025traffic} has therefore pursued wave smoothing, and a wide range of evidence, including simulations~\cite{ghiasi2019mixed,li2024jam}, controlled experiments~\cite{stern2018dissipation,stern2019quantifying}, field tests~\cite{yang2018field}, and empirical-replay studies~\cite{yang2014control,coifman2026using}, supports the conclusion that reducing these oscillations can deliver meaningful emission reductions.

Yet a fundamental question remains largely unaddressed: what is the emission reduction potential of large-scale stop-and-go wave smoothing in real-world traffic? Most prior estimates are drawn from controlled settings or limited datasets that cannot fully represent the variability and complexity of real-world freeway dynamics~\cite{ji2026scalable}. With the recent emergence of large-scale trajectory datasets~\cite{ztd2018,gloudemans202324,chaudhari2025mitra}, we can now address a policy-relevant question directly from data: what proportion of congestion-related emissions can be eliminated by smoothing the stop-and-go waves themselves?

\begin{figure}[htpb]
    \centering
    \includegraphics[width=0.85\linewidth]{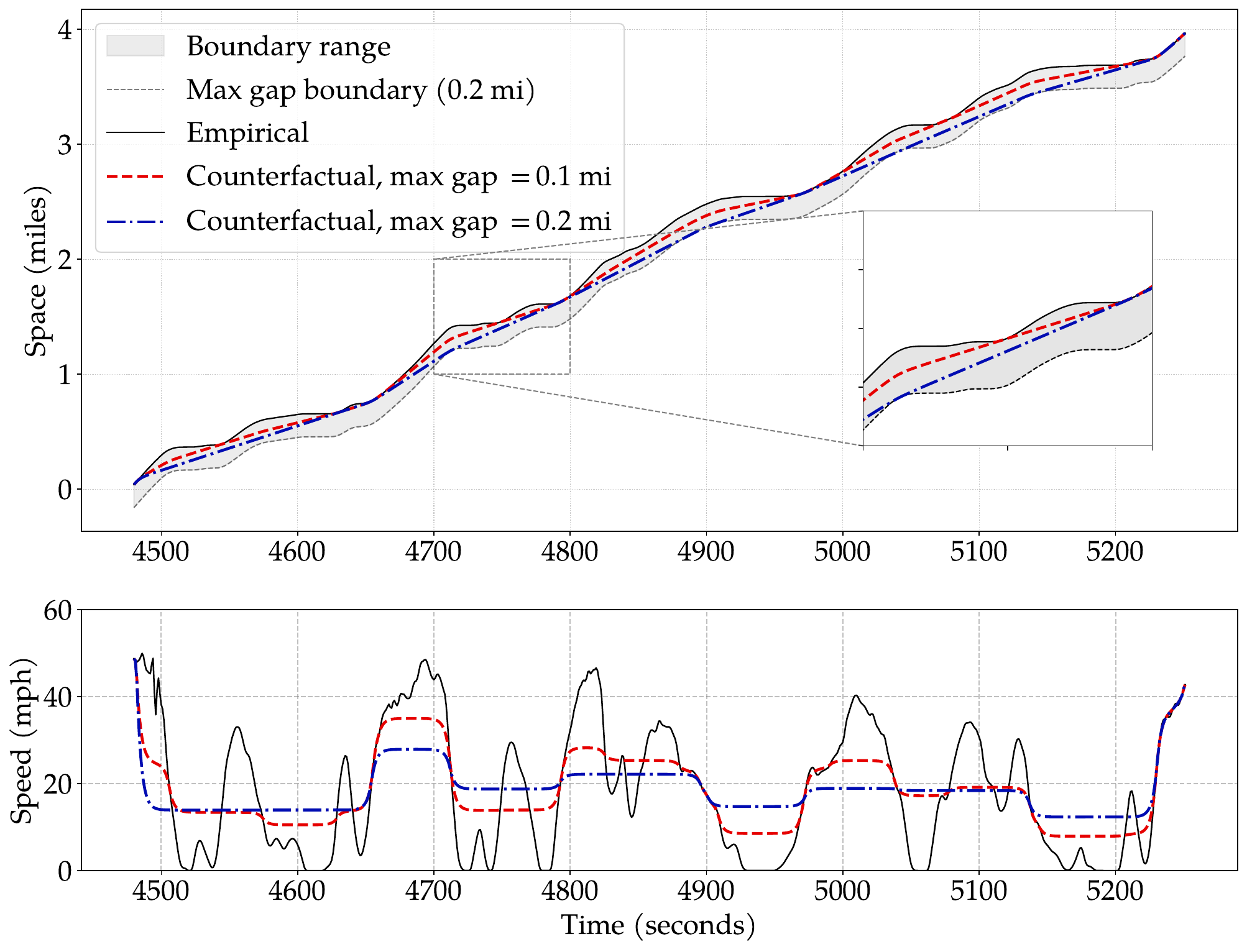} 
    \caption{
        \textbf{Comparison using the counterfactual wave-smoothing benchmark.} An example empirical trajectory from June 18, 2024, in Lane~1 is shown together with its benchmark counterparts under 0.1-mile and 0.2-mile maximum-gap constraints. (Top) Time--space diagram showing the original empirical trajectory and the benchmark trajectories constrained within the boundary range, with an inset magnifying one wave. (Bottom) Speed profiles highlighting the elimination of stop-and-go waves from the benchmark trajectories while preserving the boundary conditions.
        }
    \label{fig:motivation}
\end{figure}
To answer this question, we define a counterfactual wave-smoothing benchmark, as illustrated in \figurename~\ref{fig:motivation}. Given a trajectory with stop-and-go dynamics, we ask: to what extent emissions can be reduced by smoothing the speed profile, subject to (i) no unrealistically large increase in headway and (ii) fixed total travel time, travel distance, and boundary conditions. 

We formulate this counterfactual benchmark as a constrained quadratic programming (QP) problem. The benchmark isolates the emissions attributable to unnecessary stop-and-go waves and provides an empirical estimate of their contribution to real-world freeway emissions. We then quantify the emission difference by estimating the emissions from both the original and benchmark trajectories using the U.S. EPA MOVES model~\cite{USEPA_Motor_Vehicle_Emission_2024}, a widely used project-level emission estimation tool in the United States. We apply this framework to more than two months of daily high-resolution speed fields from the I-24 MOTION testbed~\cite{gloudemans202324,ji2025wavex}, which captures rich stop-and-go wave dynamics~\cite{ji2026scalable} in real-world freeway traffic. From these speed fields, we reconstruct virtual trajectories~\cite{tsanakas2022generating,ji2024virtual}; that is, \textbf{the trajectories analyzed in this study are reconstructions from measured speed fields rather than directly observed raw vehicle trajectories}. The diversity and scale of this dataset in terms of stop-and-go waves allow us to provide insights and discuss the implications of our findings for traffic management and control strategies aimed at stop-and-go wave mitigation.

The contributions of this study are as follows:
\begin{enumerate}[label=(\roman*),noitemsep, parsep=1pt, topsep=1pt, partopsep=0.5pt]
    \item We present the first study to benchmark the emission reduction potential of freeway stop-and-go wave smoothing using large-scale, high-resolution empirical trajectory data;
    \item We are the first to incorporate practical headway constraints into wave-smoothing benchmarks, yielding more realistic estimates of emission reduction potential;
    \item We quantify the emission reduction potential across vehicle classes and identify heavy-duty trucks as particularly attractive targets for wave smoothing, with roughly two to three times the reduction potential of light-duty vehicles and a concentration in the lanes where the potential is highest;
    \item We provide planning tools and actionable insights for traffic management and control strategies aimed at mitigating stop-and-go waves and reducing emissions.
\end{enumerate}

The remainder of this paper is organized as follows. 
\sectionname~\ref{sec:related} reviews related work on wave-smoothing strategies and their reported benefits. \sectionname~\ref{sec:method} presents the methodology for benchmarking the emission reduction potential of stop-and-go wave smoothing using large-scale trajectory datasets. \sectionname~\ref{sec:data} describes the dataset used in this study. \sectionname~\ref{sec:results} presents the main results. \sectionname~\ref{sec:discussion} discusses the implications of our findings for practitioners. \sectionname~\ref{sec:conclusion} concludes the paper and outlines future work.

\section{Related work}
\label{sec:related}
Emission reduction in road traffic can be pursued through three primary pathways: capacity-based strategies (expanding roadway capacity), demand-based strategies (reducing travel demand), and vehicle-based strategies (changing driving behavior)~\cite{bigazzi2012congestion}. Among these, vehicle-based approaches often provide more significant and directly controllable emission reductions~\cite{xia2023future}, particularly under congested conditions~\cite{barth2008real,bigazzi2012congestion,bigazzi2017can}. Interest in vehicle-based strategies has also grown rapidly with the rise of connected and automated vehicles (CAVs) and modern intelligent transportation systems (ITS), which make fine-grained, real-time control increasingly feasible.

Recent reviews reflect this momentum.
Jiang et al.~\cite{jiang2025environmental} provide a comprehensive review of vehicle-based emission reduction strategies, covering a broad range of approaches, such as automated vehicles at different levels of automation, variable speed limits, and eco-routing. Vahidi and Sciarretta~\cite{vahidi2018energy} examine the energy and emission benefits of CAVs in scenarios that emphasize anticipation and cooperation, with particular attention to platoon and signal control. However, neither review focuses specifically on mitigating stop-and-go traffic waves as an operational mechanism for reducing emissions. We therefore provide a targeted discussion of wave-smoothing methods and their documented environmental impacts.

\subsection{Stop-and-go wave smoothing strategies}
\label{sec:strategies}

A pioneering study by Li et al.~\cite{li2014stop} analyzed empirical NGSIM trajectory data and identified a strong correlation between stop-and-go wave amplitude and vehicle emissions, establishing an early quantitative link between stop-and-go traffic and increased emissions. Building on this insight, researchers have explored a wide range of mitigation strategies through numerical simulations, empirical-data replay, closed-track experiments, and small-scale field tests. These approaches, variously termed stop-and-go wave smoothing or jam-absorbing driving (JAD)~\cite{nishi2013theory,li2024jam}, speed harmonization~\cite{ma2016freeway}, traffic oscillation mitigation~\cite{li2014stop}, stop-and-go wave dissipation~\cite{stern2018dissipation,vcivcic2019stop,goatin2024dissipation}, and stop-and-go wave suppression~\cite{he2026review}, share a common control objective. Specifically, when stop-and-go waves are present, vehicles proactively reduce speed and increase headway to attenuate traffic disturbances, thereby minimizing unnecessary acceleration and deceleration and ultimately smoothing traffic flow.

It is worth noting that a separate line of research focuses on developing string-stable controllers~\cite{ploeg2013controller,wang2018infrastructure,gunter2020commercially,bahavarnia2025constrained}, which, by definition, suppress traffic instabilities and thereby prevent stop-and-go waves from forming. This line of work is beyond the scope of this study. A further consideration is lateral dynamics. Because stop-and-go waves emerge and propagate longitudinally, the strategies reviewed above rely on longitudinal speed control~\cite{stern2018dissipation,vishnoi2024cav,lee2025traffic}; a recent dedicated review~\cite{he2026review} surveys the state of practice in this area. Lateral maneuvers, particularly lane changes, are nonetheless well documented as triggers and amplifiers of freeway oscillations~\cite{ahn2007freeway,laval2010mechanism,zheng2011freeway}. Control strategies that coordinate lateral behavior are beyond the scope of this study.

An early precursor is the freeway-based dynamic eco-driving system of Barth and Boriboonsomsin~\cite{barth2009energy}, which provides real-time speed advice based on prevailing traffic conditions and reports 10--20\% reductions in fuel consumption and \ce{CO2} emissions under congested conditions.
Yang and Jin~\cite{yang2014control} propose a wave-smoothing controller based on Newell's car-following model and evaluate it using NGSIM data, reporting up to a 34\% reduction in simulated \ce{CO2} emissions. Yang et al.~\cite{yang2018field} further validate this approach in a small-scale proof-of-concept field test on urban roads under mild congestion, in which a downstream vehicle provides speed and position information to the controlled vehicle.
Stern et al.~\cite{stern2018dissipation} introduce the FollowerStopper controller, which uses information from the leading vehicle as feedback to dissipate waves, and test the method on a closed track.
Wu et al.~\cite{wu2021flow} subsequently replicate the experiment in simulation and design reinforcement-learning (RL) controllers for wave mitigation. RL-based approaches have since proliferated; for example, Jang et al.~\cite{jang2025reinforcement} test and further develop such algorithms using empirical data collected from real-world freeway traffic.
Another line of research concerns jam-absorbing driving, which originated in a citizen-science blog~\cite{beaty1998traffic} and was later formalized by Nishi et al.~\cite{nishi2013theory} based on the ``slow-in, fast-out'' principle, establishing how a single vehicle can absorb a traffic jam. Li et al.~\cite{li2024jam} further develop this line of work by estimating wave propagation and testing the approach in microsimulation.
In a recent study, Coifman and Liu~\cite{coifman2026using} propose a simple yet effective wave-smoothing controller based on Newell's model. The controller leverages downstream platoon information, propagates it at a fixed wave speed, and stitches trajectories together to plan speeds in a rolling-horizon framework. Their results demonstrate the strong performance of this relatively simple and robust method, which inspires the approach used in this study.

Regardless of the specific control algorithm, wave smoothing is typically achieved by attenuating speed fluctuations through the creation of additional headway: the controlled vehicle decelerates earlier and maintains an enlarged gap, thereby forming a buffer that absorbs upstream traffic waves. In practice, however, the required gap can be substantial (up to 0.2 miles, or approximately 320 meters)~\cite{li2024jam} and is often impractical in real-world freeway operations. Such large gaps may raise safety and operational concerns by (\romannumeral1) encouraging other vehicles to enter the gap and (\romannumeral2) provoking tailgating by impatient following drivers. Both behaviors were observed in prior field experiments~\cite{matt2024middle}, indicating an inherent trade-off between wave-smoothing effectiveness and feasible headway maintenance.

Despite these practical limitations, existing studies largely assume that large gaps can be sustained, potentially leading to an overestimation of the achievable emission-reduction benefits of wave-smoothing strategies. A systematic analysis of the trade-off between wave-smoothing benefits and the additional gap required is therefore needed to bridge the divide between theory and practice and to provide freeway operations practitioners with a more realistic assessment of emission-reduction potential.

\begin{landscape} 
\begin{table}[htpb]
\centering
\small
\setlength{\tabcolsep}{6pt}
\renewcommand{\arraystretch}{1.15}
\caption{\textbf{Summary of representative stop-and-go wave-smoothing methods in the literature.} The methods are compared in terms of their control input, empirical data source, field testing, control scope (S: single vehicle; M: multiple vehicles), maximum-gap constraint, emission model, and reported \ce{CO2} reduction. Note that a reduction in fuel consumption is equivalent to a reduction in \ce{CO2} emissions. The commonly used emission models are CMEM~\cite{barth2001recent}, MOVES~\cite{USEPA_Motor_Vehicle_Emission_2024}, VT-Micro~\cite{rakha2004development}, SIDRA~\cite{akccelik2014recalibration}, and CIRCLES~\cite{khoudari2023reducing}.}
\label{tab:wave-smoothing-methods}
\begin{tabular*}{\linewidth}{@{\extracolsep{\fill}} c l l c c c c c c @{}}
\toprule
& \textbf{Method} 
& \textbf{Control Input}
& \makecell{\textbf{Data} \\ \textbf{Source}} 
& \makecell{\textbf{Field} \\ \textbf{Test}} 
& \makecell{\textbf{Control} \\ \textbf{Scope}} 
& \makecell{\textbf{Max. Gap} \\ \textbf{Constraint}} 
& \makecell{\textbf{Emission} \\ \textbf{Model}} 
& \makecell{\textbf{CO}$_2$ \\ \textbf{Reduction}}  \\
\midrule
\cite{yang2014control} & Newell's model &  Downstream vehicle & NGSIM  & \xmark  & S &\xmark & CMEM & 34\% \\
\cite{yang2018field}  &  Newell's model & Downstream vehicle &  Field data    & \checkmark & S &\xmark & CMEM & 20\% \\
\cite{stern2018dissipation} & FollowerStopper & Leading vehicle & Closed track  & \checkmark & S & \xmark & MOVES & 15\% \\

\cite{ghiasi2019mixed} & Newell's model & Downstream sensor & Simulation  &\xmark & S &\xmark & VT-micro & 2--29\% \\
\cite{li2024jam} & JAD theory &  Downstream platoon & Simulation  &\xmark  & S &\xmark & SIDRA& 10\% \\
\cite{jang2025reinforcement} & RL-based  & Downstream speed & I-24 Drive  & \xmark & M &\xmark &CIRCLES & 35\% \\
\cite{jang2025reinforcement} & RL-based  & Downstream speed & Field data  & \checkmark & M & \xmark &CIRCLES& 12--16\% \\
\cite{coifman2026using} & Newell's model &Downstream platoon & NGSIM+  & \xmark & S, M & \xmark &CMEM& 20--35\% \\
\bottomrule
\end{tabular*}
\end{table}
\end{landscape}
\subsection{Emission reductions via wave smoothing}
Prior studies report a wide range of potential \ce{CO2} emission reductions, from 2\% to 40\% (see \tablename~\ref{tab:wave-smoothing-methods}). This variability is largely attributable to differences in (\romannumeral 1) emission estimation methods, (\romannumeral 2) data sources and scenarios, and (\romannumeral 3) control strategies, as discussed in \sectionname~\ref{sec:strategies}. The reported emission reduction potential depends strongly on the emission model employed because not all pollutants are consistently quantified across studies. For example, studies using VT-Micro~\cite{rakha2004development} often focus primarily on fuel consumption and \ce{CO2} emissions and provide limited information about other pollutants. In contrast, the U.S. EPA MOVES model~\cite{USEPA_Motor_Vehicle_Emission_2024} provides a more comprehensive representation of vehicular emissions and is the standard tool for transportation conformity analyzes in the United States, as mandated by~\cite{EPA2024_MOVES5_Notice_2024_29073}. However, owing to its complexity at the project level~\cite{ramirez2025neuralmoves}, it has seen limited adoption in prior wave-smoothing studies~\cite{stern2019quantifying}. Data sources likewise differ substantially in their ability to capture stop-and-go traffic dynamics. In the well-known ring-road experiments~\cite{stern2018dissipation,stern2019quantifying}, all vehicles operated at speeds below 30 mph, and the observed wave intensity (about 7 waves per mile) was roughly an order of magnitude higher than that typically observed under real-world conditions~\cite{ji2026scalable}. Similarly, the NGSIM dataset contains only about seven waves in total, which is insufficient to capture the full range of stop-and-go dynamics. Although field tests are more realistic, they often involve a limited number of vehicles, and traffic dynamics can vary considerably with the day and time of testing. Together, these discrepancies highlight the need for a standardized benchmarking framework that enables consistent and comparable evaluations of wave-smoothing strategies.
\section{Methodology}
\label{sec:method}

\begin{figure}[htpb]
    \centering
    \includegraphics[width=\linewidth]{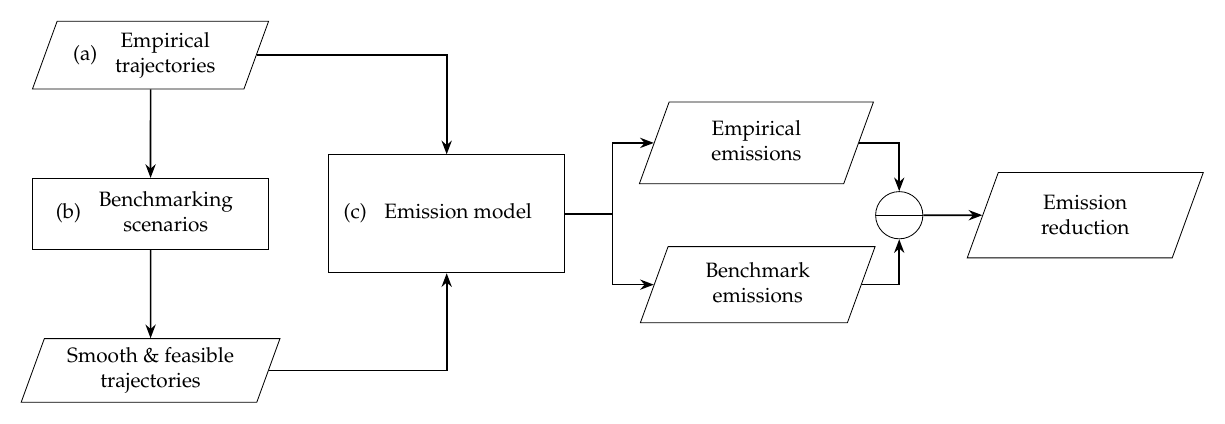} 
    \caption{
        \textbf{Overview of the framework for benchmarking the emission costs of freeway stop-and-go waves.} (a) Empirical vehicle trajectories from the I-24 MOTION testbed~\cite{gloudemans202324} provide more than two months of high-resolution data capturing rich stop-and-go wave dynamics. (b) Benchmarking scenarios are defined, and smooth, feasible trajectories are reconstructed. (c) An emission model is expressed as a function of speed and acceleration; we implement the U.S. EPA MOVES model~\cite{USEPA_Motor_Vehicle_Emission_2024} to obtain project-level estimates.
        }
    \label{fig:overview}
\end{figure}

To measure the emission costs of stop-and-go waves, we propose a benchmarking framework consisting of three key components, as illustrated in \figurename~\ref{fig:overview}: (a) empirical vehicle trajectories from the I-24 MOTION testbed~\cite{gloudemans202324}, which provide more than two months of high-resolution data capturing rich stop-and-go wave dynamics; (b) benchmarking scenarios in which smooth, feasible trajectories are reconstructed under practical constraints; and (c) an emission model expressed as a function of speed $v$ and acceleration $a$, for which we implement the U.S. EPA MOVES model~\cite{USEPA_Motor_Vehicle_Emission_2024} to obtain project-level estimates. Each component is discussed in detail below.

\subsection{Empirical trajectories}
The empirical trajectories in this study are generated from high-resolution mean speed fields constructed using Edie's definition~\cite{edie1963discussion}, as detailed in~\cite{ji2024virtual,ji2026scalable}. The trajectories are obtained by solving the following ordinary differential equation (ODE) over the mean speed field $v_S(t,x)$~\cite{tsanakas2022generating}, which has a spatiotemporal resolution of 0.02 miles and 4 seconds~\cite{ji2026scalable}:

\begin{equation}
\frac{\mathrm{d}x(t)}{\mathrm{d}t} = v_S(t,x(t)),
\label{eq:euler}
\end{equation}

given the initial condition $x(t=0) = x_0$, where $x(t)$ is the vehicle position at time $t$ and $v_S(t,x)$ is the mean speed field at time $t$ and position $x$. Following~\cite{tsanakas2022generating}, we apply bicubic interpolation to the speed field to improve the accuracy of the ODE solution.

\subsection{Benchmarking smooth and feasible trajectories}

Extensive research has shown that dissipating traffic waves typically requires vehicles to increase headway and reduce speed fluctuations~\cite{stern2018dissipation,yang2014control,li2014stop}. However, excessively large gaps can trigger more cut-ins or induce aggressive following from behind, creating unsafe conditions~\cite{matt2024middle}. Motivated by this trade-off, we construct a smooth and feasible benchmark trajectory from each empirical trajectory by solving a quadratic program (QP) with practical constraints.

Consider an empirical vehicle trajectory sampled over a fixed horizon of $N$ intervals with step size $\Delta t$ and total duration $T=N\Delta t$,

\begin{equation}
\mathbf{x}_{\mathrm{ref}}=[x_{\mathrm{ref},0},\dots,x_{\mathrm{ref},N}]^\top\in\mathbb{R}^{N+1}.
\label{eq:xref}
\end{equation}

From the empirical trajectory in \eqref{eq:xref}, we obtain the initial and terminal speed/acceleration values $(v_{\text{start}},a_{\text{start}})$ and $(v_{\text{end}},a_{\text{end}})$, computed consistently with the discrete operators defined in~\eqref{eq:dynamics}. We seek a benchmark trajectory $\mathbf{x}=[x_0,\dots,x_N]^\top$ that shares the same endpoints, travel time, and endpoint derivatives as the empirical trajectory. These conditions ensure that any estimated emission difference is attributable to reduced speed oscillations rather than to changes in trip length, duration, or boundary conditions.

A naive constant-speed interpolation between $(0,x_{\mathrm{ref},0})$ and $(T,x_{\mathrm{ref},N})$ is generally infeasible under congested traffic conditions. Because the empirical vehicle exhibits stop-and-go dynamics, a constant-speed profile may ``overtake'' the empirical trajectory in the time--space diagram, which is physically unrealistic. Conversely, allowing the benchmark to fall too far behind $\mathbf{x}_{\mathrm{ref}}$ corresponds to opening an excessively large headway, which is unsafe in real-world operation because it invites cut-ins and close following~\cite{matt2024middle}. We therefore impose two constraints: (\romannumeral1) no overpass, requiring the benchmark trajectory never to exceed the empirical one, and (\romannumeral2) maximum gap, requiring the benchmark to remain within a maximum distance $\delta_{\text{gap}}$ behind the empirical trajectory.

Let $\mathbf{D}_1\in\mathbb{R}^{N\times(N+1)}$, $\mathbf{D}_2\in\mathbb{R}^{(N-1)\times(N+1)}$, and $\mathbf{D}_3\in\mathbb{R}^{(N-2)\times(N+1)}$ denote the first-, second-, and third-order difference matrices such that the discrete velocity, acceleration, and jerk sequences are
\begin{equation}
\mathbf{v}=\mathbf{D}_1\mathbf{x},\qquad \mathbf{a}=\mathbf{D}_2\mathbf{x},\qquad \mathbf{j}=\mathbf{D}_3\mathbf{x},
\label{eq:dynamics}
\end{equation}
equivalently,
\begin{equation}
\begin{aligned}
v_k &:= \frac{x_{k+1}-x_k}{\Delta t}, && k=0,\dots,N-1,\\
a_k &:= \frac{x_{k+2}-2x_{k+1}+x_k}{\Delta t^2}, && k=0,\dots,N-2,\\
j_k &:= \frac{x_{k+3}-3x_{k+2}+3x_{k+1}-x_k}{\Delta t^3}, && k=0,\dots,N-3.
\end{aligned}
\label{eq:componentwise}
\end{equation}

In addition to the two gap constraints, we bound the sequences defined in \eqref{eq:componentwise} so that the benchmark represents a physically executable vehicle motion rather than a merely smooth curve: the vehicle never reverses, and its deceleration, acceleration, and jerk remain within the limits $a_{\text{dec}}^{\max}>0$, $a_{\text{acc}}^{\max}>0$, and $j^{\max}>0$, respectively.

We choose the benchmark by penalizing deviations from the constant average speed
\begin{equation}
\bar v=\frac{x_{\mathrm{ref},N}-x_{\mathrm{ref},0}}{T},
\label{eq:vbar}
\end{equation}
and nonsmooth acceleration, since both speed variability and aggressive transients are associated with higher emissions. The resulting QP optimization problem is
\begin{equation}
\begin{aligned}
    \min_{\mathbf{x}\in\mathbb{R}^{N+1}}\quad
    & \underbrace{\left\|\mathbf{D}_1\mathbf{x}-\bar v\mathbf{1}\right\|_2^2}_{\text{deviation from average speed}} &
    \;+\;
    \lambda\underbrace{\left\|\mathbf{D}_2\mathbf{x}\right\|_2^2}_{\text{acceleration smoothness}} \\
    \text{s.t.}\quad
    &\text{(boundary conditions)}& x_0 = x_{\text{ref},0},\\
    & & x_N = x_{\text{ref},N},\\
    & & (\mathbf{D}_1\mathbf{x})_0 = v_{\text{start}},\\
    & & (\mathbf{D}_1\mathbf{x})_{N-1} = v_{\text{end}},\\
    &\text{(no overpass)} & \mathbf{x} \preceq \mathbf{x}_{\text{ref}},\\
    &\text{(maximum gap)} & \mathbf{x} \succeq \mathbf{x}_{\text{ref}} - \delta_{\text{gap}}\mathbf{1} ,\\
    &\text{(no reversing)} & \mathbf{D}_1\mathbf{x} \succeq \mathbf{0},\\
    &\text{(deceleration limit)} & \mathbf{D}_2\mathbf{x} \succeq -a_{\text{dec}}^{\max}\mathbf{1},\\
    &\text{(acceleration limit)} & \mathbf{D}_2\mathbf{x} \preceq a_{\text{acc}}^{\max}\mathbf{1},\\
    &\text{(jerk limits)} & -j^{\max}\mathbf{1}\preceq\mathbf{D}_3\mathbf{x} \preceq j^{\max}\mathbf{1}.
\end{aligned}
\label{eq:QP}
\end{equation}
Here $\lambda\ge0$ is a regularization parameter for acceleration smoothness, and $\delta_{\text{gap}}\ge0$ specifies the maximum gap of the benchmark relative to the empirical trajectory. Because the initial and final positions are fixed, the sum of the discrete velocities is also fixed; consequently, $\|\mathbf{D}_1\mathbf{x}-\bar v\mathbf{1}\|_2^2$ and $\|\mathbf{D}_1\mathbf{x}\|_2^2$ differ only by a constant independent of $\mathbf{x}$. We present the former in \eqref{eq:QP} because it expresses more directly the objective of reducing speed variability around the average speed in \eqref{eq:vbar}.

When endpoint acceleration continuity is also required, the boundary conditions in \eqref{eq:QP} are augmented with
\begin{equation}
(\mathbf{D}_2\mathbf{x})_0 = a_{\text{start}}, \qquad (\mathbf{D}_2\mathbf{x})_{N-2} = a_{\text{end}}.
\label{eq:accel_continuity}
\end{equation}
The equalities in \eqref{eq:accel_continuity} may be omitted for short trajectory segments, for which empirical endpoint acceleration estimates would otherwise overconstrain the feasible set. Since the QP in \eqref{eq:QP} is convex, it can be solved efficiently to global optimality using standard solvers. The convexity also makes the approach easy to scale to large-scale datasets, as we demonstrate in \sectionname~\ref{sec:results}.

\subsection{Emission model}
The emission model used in this study is based on the U.S. EPA MOVES model~\cite{USEPA_Motor_Vehicle_Emission_2024}, which is widely used for project-level emission estimation in the United States~\cite{sentoff2015implications,stern2018dissipation,vieira2023estimating}.
It provides estimates of fuel consumption (which is equivalent to carbon dioxide, \ce{CO2}, in MOVES) and emissions of various pollutants, including carbon monoxide (CO), hydrocarbons (HC), and nitrogen oxides (\ce{NOx}).
Although MOVES is a standard tool for project-level emission estimation, its complexity and data requirements can make it challenging for researchers outside the transportation emission research field to implement. Consequently, it is less commonly used than VT-Micro~\cite{rakha2004development} and CMEM~\cite{barth2001recent}. To clarify its application in this study, we provide a brief overview of the MOVES model.

The MOVES model estimates emission rates by engine operating mode, which is defined by the combination of vehicle speed and vehicle-specific power (VSP)~\cite{jimenez1998understanding,epa2024moves5activity}. VSP is calculated as follows:

\begin{equation}
VSP = \frac{Av+Bv^2+Cv^3+M_s(a+g \cdot \sin\theta)\cdot v}{M_f},
\label{eq:vsp}
\end{equation}

where $A$, $B$, and $C$ are the road-load coefficients in units of $kW \cdot s/m$, $kW \cdot s^2/m^2$, and $kW \cdot s^3/m^3$, respectively. $M_s$ is the source vehicle mass in tons, and $M_f$ is the fixed mass factor in tons. Furthermore, $g$ is gravitational acceleration ($9.8~m/s^2$), $\theta$ is the road grade in radians, $v$ is the instantaneous speed in $m/s$, and $a$ is the instantaneous acceleration in $m/s^2$. 

\begin{figure}[htpb] 
\centering 
\begin{tikzpicture} 
\begin{axis}[ 
    width=0.9\linewidth, 
    height=0.7\linewidth, 
    xlabel={Speed $v$ (mph)}, 
    ylabel={VSP (kW/tonne)}, 
    xmin=0, xmax=80, 
    ymin=-6, ymax=36, 
    grid=none,
    ytick={0, 3, 6, 9, 12, 18, 24, 30},
    xtick={0, 25, 50},
    extra x ticks={80},
    extra x tick labels={+},
    extra y ticks={-6, 36},
    extra y tick labels={-,+},
    axis background/.style={fill=gray!5}
] 

\draw[fill=red!10] (1,-6) rectangle (25,0) node[midway] {\small \opid{11}};
\draw[fill=red!10] (25,-6) rectangle (50,0) node[midway] {\small \opid{21}};

\draw[fill=blue!10] (1,0)  rectangle (25,3)  node[midway] {\small \opid{12}};
\draw[fill=blue!10] (1,3)  rectangle (25,6)  node[midway] {\small \opid{13}};
\draw[fill=blue!10] (1,6)  rectangle (25,9)  node[midway] {\small \opid{14}};
\draw[fill=blue!10] (1,9)  rectangle (25,12) node[midway] {\small \opid{15}};
\draw[fill=blue!10] (1,12) rectangle (25,36) node[midway] {\small \opid{16}};

\draw[fill=green!10] (25,0)  rectangle (50,3)  node[midway] {\small \opid{22}};
\draw[fill=green!10] (25,3)  rectangle (50,6)  node[midway] {\small \opid{23}};
\draw[fill=green!10] (25,6)  rectangle (50,9)  node[midway] {\small \opid{24}};
\draw[fill=green!10] (25,9)  rectangle (50,12) node[midway] {\small \opid{25}};
\draw[fill=green!10] (25,12) rectangle (50,18) node[midway] {\small \opid{27}};
\draw[fill=green!10] (25,18) rectangle (50,24) node[midway] {\small \opid{28}};
\draw[fill=green!10] (25,24) rectangle (50,30) node[midway] {\small \opid{29}};
\draw[fill=green!10] (25,30) rectangle (50,36) node[midway] {\small \opid{30}};

\draw[fill=orange!10] (50,0)  rectangle (80,6)  node[midway] {\small \opid{33}};
\draw[fill=orange!10] (50,6)  rectangle (80,12) node[midway] {\small \opid{35}};
\draw[fill=orange!10] (50,12) rectangle (80,18) node[midway] {\small \opid{37}};
\draw[fill=orange!10] (50,18) rectangle (80,24) node[midway] {\small \opid{38}};
\draw[fill=orange!10] (50,24) rectangle (80,30) node[midway] {\small \opid{39}};
\draw[fill=orange!10] (50,30) rectangle (80,36) node[midway] {\small \opid{40}};

\end{axis} 
\end{tikzpicture} 
\caption{\textbf{Operating Mode ID (OpModeID) map defined by speed and VSP in the latest U.S. EPA MOVES5 model~\cite{USEPA_Motor_Vehicle_Emission_2024}, excluding braking (\opid{0}) and idling (\opid{1}).} Each rectangle represents an operating mode with a unique ID that corresponds to a specific emission rate for each pollutant.}
\label{fig:opmodeid-map}
\end{figure}

As shown in \figurename~\ref{fig:opmodeid-map}, operating modes are categorized into bins based on speed and the VSP defined in \eqref{eq:vsp}, with each bin assigned a unique Operating Mode ID (OpModeID). Each OpModeID corresponds to pollutant-specific emission rates derived from extensive empirical testing and modeling by the EPA. The emissions from a vehicle trajectory are estimated as
\begin{equation}
\text{Emission} = \sum_{k=0}^{N-1} \text{ER}(\text{OpModeID}(v_k, VSP_k(v_k, a_k))) \cdot \Delta t,
\label{eq:emission}
\end{equation}
where $\text{ER}(\cdot)$ is the pollutant-specific emission-rate function provided by the MOVES model. Using \eqref{eq:emission}, we compute the emissions from both the empirical and benchmark trajectories and compare the results to quantify the emission costs of stop-and-go waves. The official \href{https://github.com/USEPA/EPA_MOVES_Model/blob/25dc6c833dd8c88198f82cee93ca30be1456df8b/generators/baserategenerator/baserategenerator.go#L1861}{MOVES source code (written in Go)} is available in the EPA repository. Following this source code, we re-implement the relevant operating-mode logic in Python. In this study, we estimate emissions of four major pollutants: \ce{CO2}, \ce{CO}, \ce{HC}, and \ce{NOx}, with five different vehicle classes (the vehicle parameter configuration can be found in \tablename~\ref{tab:classspec}). We report the results mainly on the passenger car and include the discussion for the other vehicle classes in \sectionname~\ref{sec:discussion}.

\begin{table}[H]
\centering
\small
\setlength{\tabcolsep}{4pt}
\caption{MOVES configuration of the vehicle classes evaluated, read
from \texttt{sourceUseTypePhysics} ($A$, $B$, $C$ in MOVES units, $B$ and $C$
scaled by $10^{3}$; $m$, $m_f$ in tonnes). $m/m_f$ represents the relative acceleration-weighting factor in the VSP calculation, normalized to 1.00 for passenger cars.}
\label{tab:classspec}
\begin{tabular}{llcccS[table-format=1.4]S[table-format=1.3]S[table-format=1.3]S[table-format=2.3]S[table-format=2.2]S[table-format=1.2]}
\toprule
Class & Fuel & \texttt{srcType} & regCls & MY &
{$A$} & {$10^{3}B$} & {$10^{3}C$} & {$m$} & {$m_f$} & {$m/m_f$} \\
\midrule
Passenger car            & gas    & 21 & 20 & 2020 & 0.1565 & 2.002 & 0.493 & 1.479 & 1.48 & 1.00 \\
Passenger truck          & gas    & 31 & 30 & 2020 & 0.2211 & 2.838 & 0.698 & 1.867 & 1.87 & 1.00 \\
Light commercial         & gas    & 32 & 30 & 2020 & 0.2350 & 3.039 & 0.748 & 2.060 & 2.06 & 1.00 \\
Single-unit short-haul   & diesel & 52 & 46 & 2020 & 0.5965 & 0.000 & 1.603 & 13.800 & 7.00 & 1.97 \\
Combination short-haul   & diesel & 61 & 47 & 2020 & 1.5090 & 0.000 & 3.745 & 24.684 & 10.00 & 2.47 \\
\bottomrule
\end{tabular}
\end{table}

\section{Data}
\label{sec:data}
The data used in this study were collected from the I-24 MOTION testbed~\cite{gloudemans202324}, a large-scale freeway traffic observation system located along a 4.2-mile stretch of I-24 near Nashville, Tennessee. The testbed comprises 276 high-resolution cameras mounted on 40 poles, providing continuous coverage of all lanes and capturing rich stop-and-go wave dynamics in real-world freeway traffic~\cite{ji2026scalable}.

\begin{figure}[htpb]
    \centering
    \includegraphics[width=0.85\linewidth]{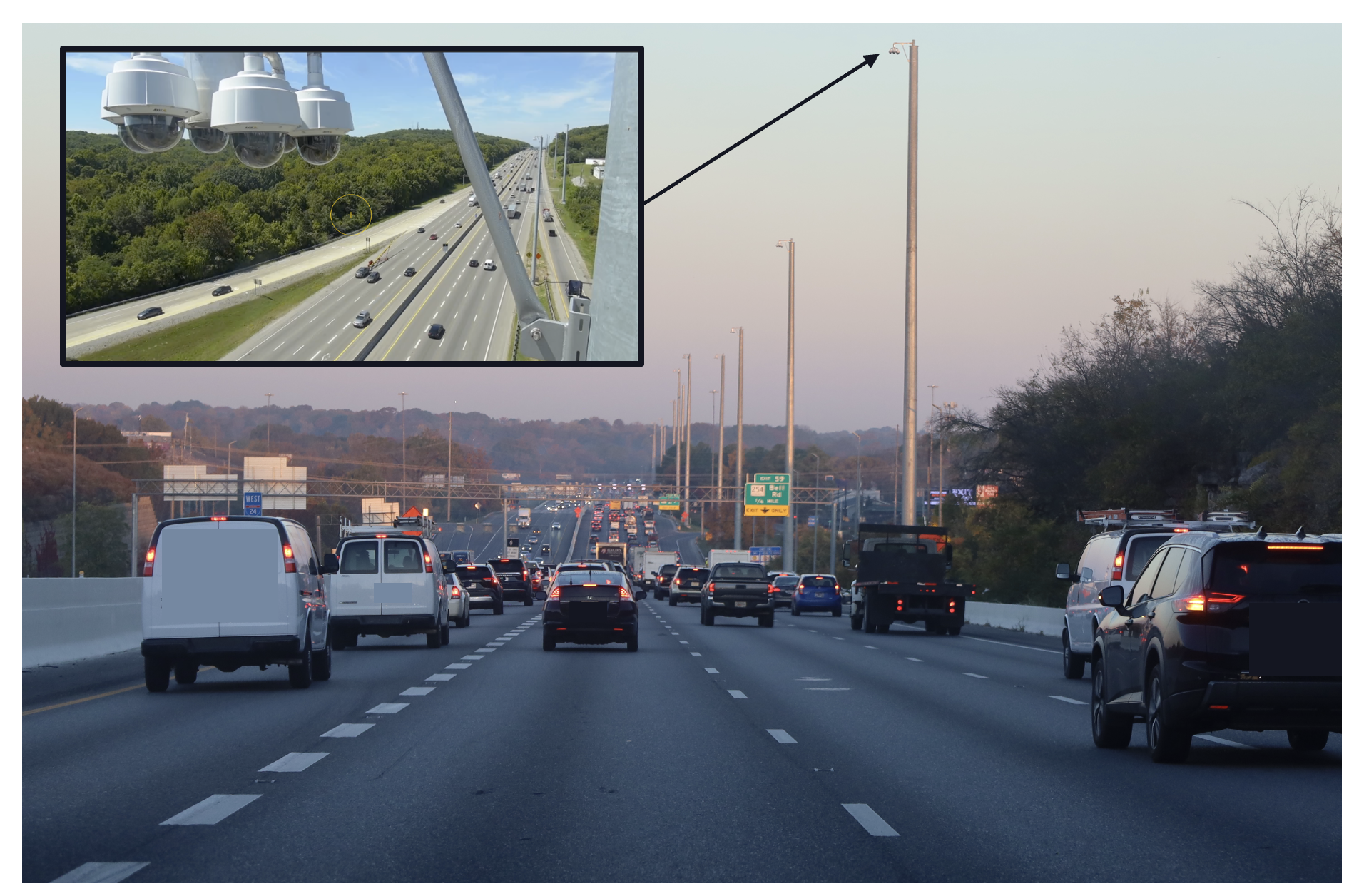} 
    \caption{\textbf{I-24 MOTION testbed.} I-24 MOTION is a large-scale freeway observation system designed to monitor traffic congestion. It consists of 276 high-resolution cameras mounted on 40 poles (as illustrated) and covers a 4.2-mile stretch of I-24 near Nashville, Tennessee.}
    \label{fig:i24-motion-map}
\end{figure}

\begin{figure}[H]
    \centering
    \includegraphics[width=0.95\linewidth]{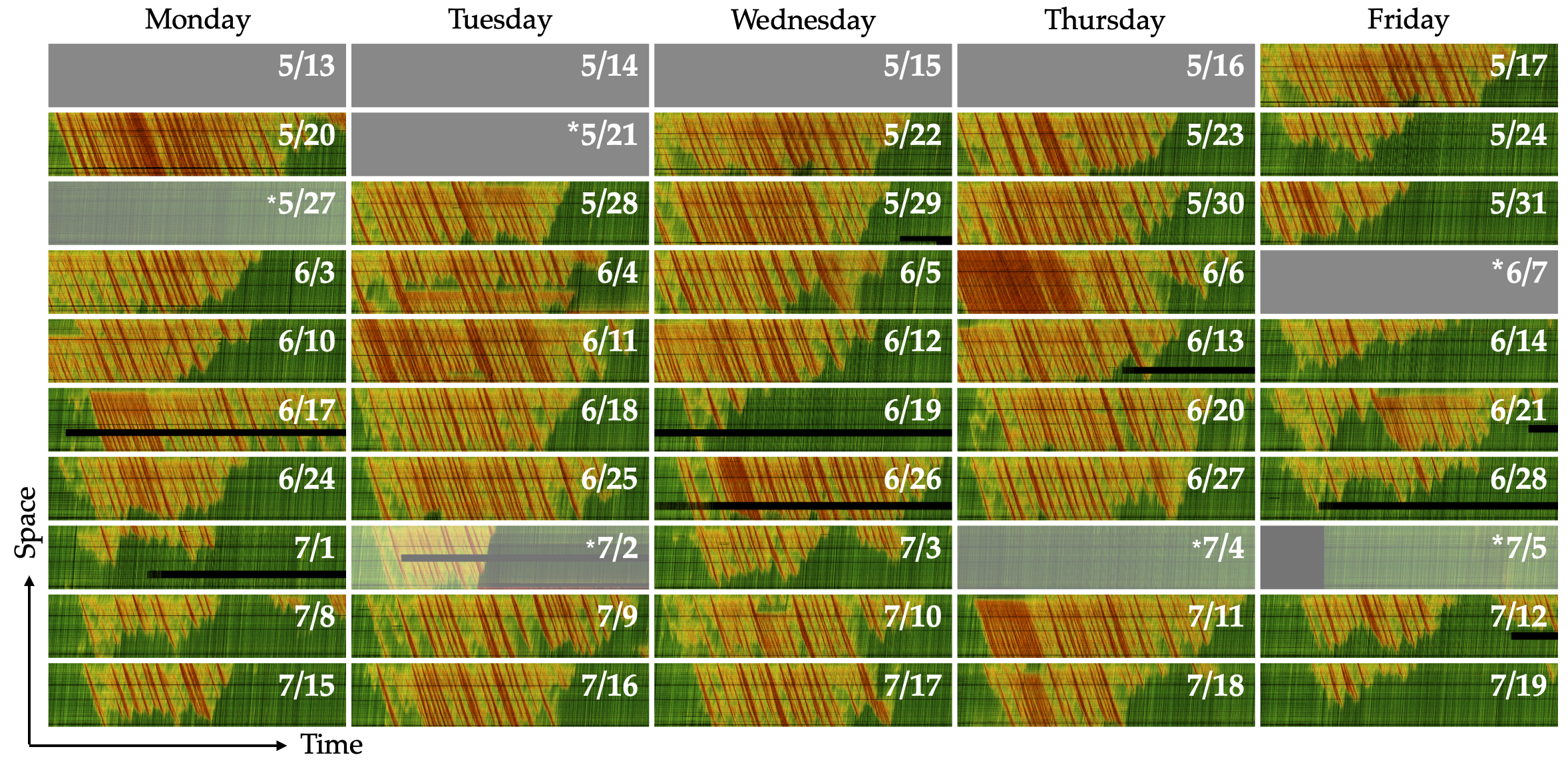} 
    \caption{\textbf{Data used in this study.} Time--space diagrams of the data collected from the I-24 MOTION testbed and analyzed in this study. The data were collected from May 17 to July 19, 2024, and cover 9 weeks of weekday peak-period traffic (6:00 a.m. to 10:00 a.m.). Each day includes 4 hours of traffic observations over the 4.2-mile testbed and exhibits rich stop-and-go wave dynamics. Holidays with light traffic and anomalous days with significant events (e.g., multiple lane closures) are excluded from the analysis and are indicated by masks and asterisks (*) next to the dates.}
    \label{fig:data-calendar}
\end{figure}

The data used in this study are visualized in \figurename~\ref{fig:data-calendar} and cover 9 weeks of weekday peak-period traffic (6:00 a.m. to 10:00 a.m.) from May 17 to July 19, 2024. To the best of our knowledge, this is the largest empirical trajectory dataset capturing stop-and-go wave dynamics in real-world freeway traffic. The daily diversity and scale of this dataset allow us to derive insights and discuss the implications of our findings for traffic management and control strategies aimed at mitigating stop-and-go waves.

Following the process documented in~\cite{ji2024stop}, we estimate the mean speed fields $v_S$ at a resolution of 4 seconds and 0.02 miles (approximately 32 meters). We then generate empirical vehicle trajectories by solving the ODE in \eqref{eq:euler}, releasing a virtual vehicle from the beginning of the testbed every 4 seconds. The sampling frequency of the virtual vehicles is set to 1 Hz to satisfy the emission model requirements. Overall, we generate more than 500,000 empirical vehicle trajectories exhibiting complex stop-and-go wave dynamics for analysis.

\section{Results}
\label{sec:results}

We apply the proposed methodology to the data described in \sectionname~\ref{sec:data} and present the results below. 
We set the QP regularization parameter to $\lambda=10$ to balance speed deviation and acceleration smoothness and conduct a sensitivity analysis of the parameter $\delta_{\text{gap}}$ over the range from 0.01 to 0.5 miles. 
For a fair comparison, we consider only vehicle trajectories with an average speed below 50 mph, as higher-speed trajectories typically do not exhibit stop-and-go dynamics and have different emission characteristics. In this section, we primarily show results from passenger car, and the average results on all day, lane 1 (the left most lane); the lane-by-lane comparison, and more vehicle classes are presented in \sectionname~\ref{sec:discussion}.

\subsection{Case study on one empirical trajectory}
\begin{figure}[H]
    \centering
    \includegraphics[width=0.95\linewidth]{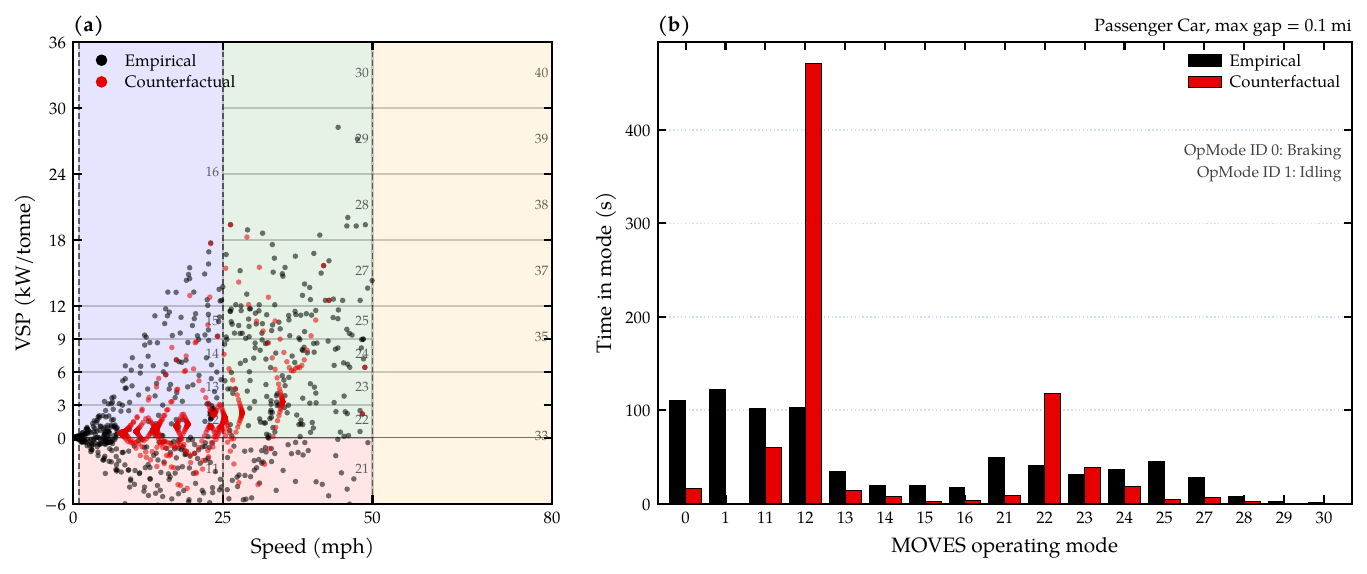}
    \caption{
        \textbf{Comparison of operating modes for the empirical and benchmark trajectories.} The example corresponds to \figurename~\ref{fig:motivation} and was sampled from Lane~1 data collected on June 18, 2024, under a 0.1-mile maximum-gap constraint.
        (a) Speed--VSP diagram showing the operating-mode distributions of the empirical and benchmark trajectories and highlighting the reduction in high-VSP modes in the benchmark scenario. 
        (b) Time spent in each MOVES operating mode by the two trajectories, illustrating the shift in the benchmark case, with substantially less braking (\opid{0}) and idling (\opid{1}).}
    \label{fig:result-summary}
\end{figure}

For clarity, we first present a detailed case study of a single vehicle trajectory sampled from Lane~1 data collected on June 18, 2024, under a 0.1-mile maximum-gap constraint. The empirical and benchmark trajectories are visualized in \figurename~\ref{fig:motivation}, which shows that the benchmark trajectory successfully smooths large stop-and-go waves while satisfying the maximum-gap constraint. The corresponding speed--VSP diagram is shown in \figurename~\ref{fig:result-summary}(a), with the empirical trajectory states shown in black and the benchmark states shown in red. The operating-mode distributions are compared in \figurename~\ref{fig:result-summary}(b); the benchmark trajectory exhibits a marked reduction in high-VSP modes and fewer instances of idling and braking. Relative to the empirical trajectory, the benchmark trajectory achieves substantial reductions in all four pollutants: 11.3\% in \ce{CO2}, 7.2\% in CO, 31.6\% in HC, and 44.1\% in \ce{NOx}. The magnitude differs markedly across pollutants; we return to the technical details behind this contrast in~\ref{sec:appendix-mechanism}.

When we relax the maximum-gap constraint to 0.2 miles, the benchmark trajectory can further smooth the speed profile, as shown in \figurename~\ref{fig:motivation}. The corresponding emission reductions increase to 13.9\% for \ce{CO2}, 8.2\% for CO, 41.4\% for HC, and 54.5\% for \ce{NOx}. Relative to the 0.1-mile case, increasing the maximum gap to 0.2 miles yields additional reductions of 2.6 percentage points for \ce{CO2}, 1.0 percentage point for CO, 9.8 percentage points for HC, and 10.4 percentage points for \ce{NOx}. Doubling the permitted gap therefore buys little additional \ce{CO2} or CO benefit, while \ce{NOx} and HC continue to improve appreciably; we examine the reason for this asymmetry in~\ref{sec:appendix-mechanism}. This case study illustrates the trade-off between the maximum gap and emission reduction, which we analyze at scale in the next section.

\subsection{Large-scale analysis for one day}
\begin{figure}[H]
    \centering
    \includegraphics[width=0.9\linewidth]{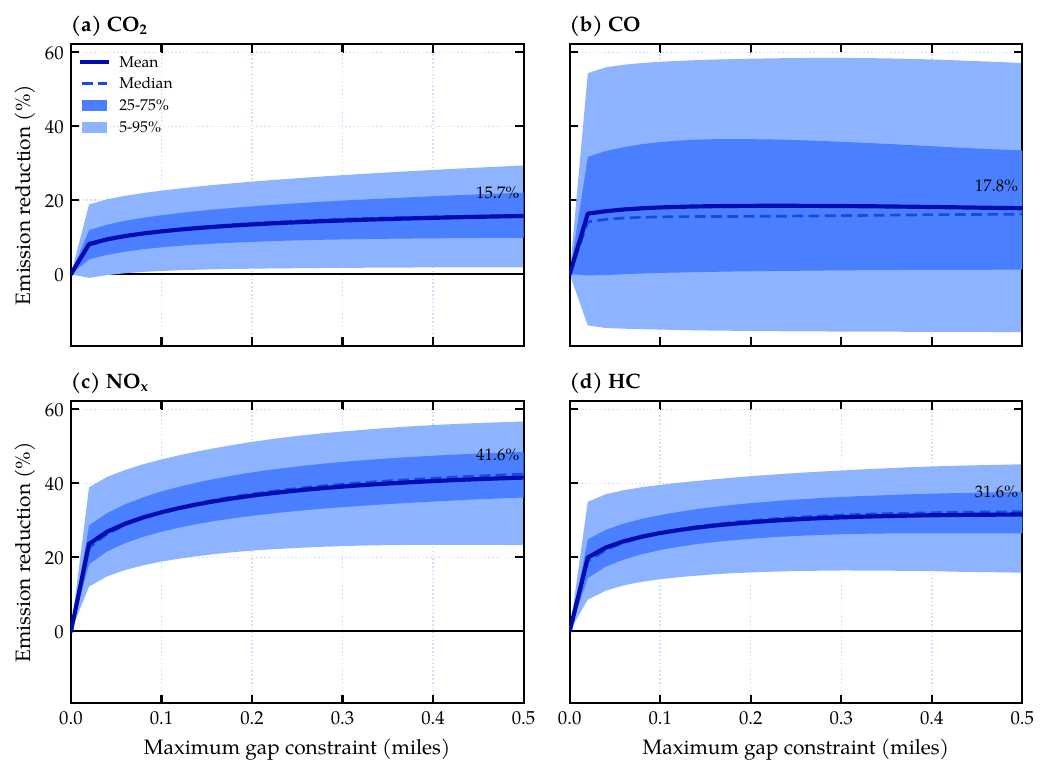}
    \caption{\textbf{Large-scale analysis for a single day (June 18, 2024) under different maximum-gap constraints.} Emission reduction percentages for (a) \ce{CO2}, (b) CO, (c) \ce{NOx}, and (d) HC are shown as functions of the maximum-gap constraint, together with the mean, median, and the 25th--75th and 5th--95th percentile bands across all empirical trajectories from that day with average speeds below 50 mph.}
    \label{fig:large-scale-analysis}
\end{figure}

We extend the single-trajectory analysis to a full day of Lane~1 data from June 18, 2024, generating benchmark trajectories under different maximum-gap constraints for trajectories with average speeds below 50 mph. \figurename~\ref{fig:large-scale-analysis} summarizes the results for that day, with distribution bounds shown for the 5th to 95th percentiles and the 25th to 75th percentiles (the interquartile range). Across different stop-and-go traffic patterns, three pollutants (\ce{CO2}, HC, and \ce{NOx}) exhibit a consistent trend: smoother traffic produces lower emissions, and allowing a larger gap produces greater reductions. CO behaves differently: its mean reduction saturates almost immediately, declines slightly for gaps beyond about 0.2 miles, and exhibits a much wider spread across trajectories, with the 5th--95th percentile band extending below zero at every gap size.
For simplicity, the remaining results in this study are reported as daily average emission reductions across all empirical trajectories.

\subsection{Day-to-day emission reduction analysis}
\begin{figure}[H]
    \centering
    \includegraphics[width=\linewidth]{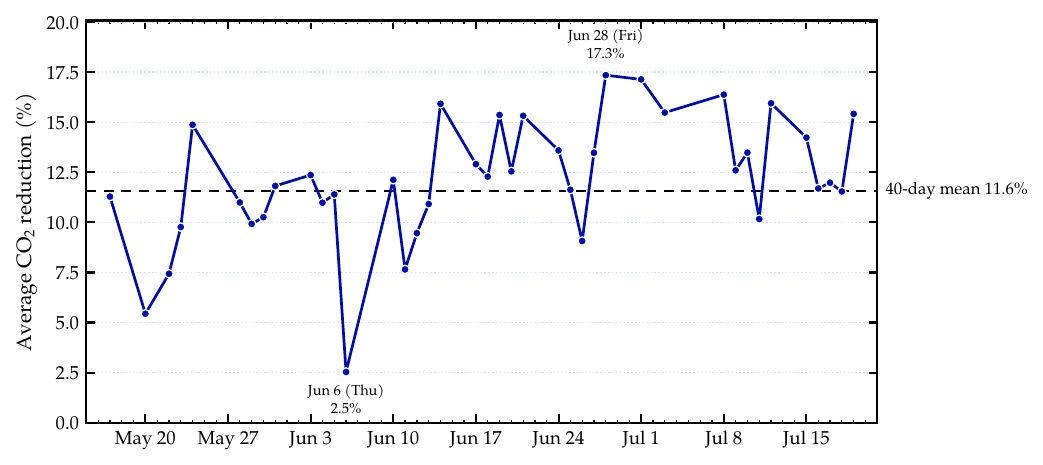}
    \caption{\textbf{Daily \ce{CO2} reduction trends from May 17 to July 19, 2024, under a 0.1-mile maximum-gap constraint.} The reduction potential varies across days, ranging from approximately 2.5\% to 17.5\%, depending on the stop-and-go wave patterns observed each day.
    }
    \label{fig:day-to-day}
\end{figure}

We further apply the proposed methodology to all data described in \sectionname~\ref{sec:data}, generating benchmark trajectories for more than 500,000 empirical trajectories under different maximum-gap constraints. To illustrate the daily trend, we use a maximum-gap constraint of 0.1 miles and summarize the average daily emission reductions in \figurename~\ref{fig:day-to-day}, which shows \ce{CO2} reduction patterns over 40 days.

The daily emission reduction patterns in~\figurename~\ref{fig:day-to-day} vary substantially, with \ce{CO2} reductions ranging from approximately 2.5\% to 17.5\% around a 40-day mean of 11.6\%. We examine several representative days to identify the factors contributing to this variability. The days with the highest reduction potential, June 28 (17.3\%), July 1 (17.1\%), and July 8 (16.4\%), 2024, exhibited relatively light traffic waves with few severe stop-and-go waves. On these days, vehicles typically encountered only one or two stop-and-go waves, and some of the reduction appears to result from smoother driving under free-flow conditions before and between these waves. 

In contrast, days with lower reduction potential tend to exhibit more intense stop-and-go waves and frequent, severe speed fluctuations. The day with the smallest reduction, June 6, 2024, had an average \ce{CO2} reduction of 2.5\% and exhibited particularly intense stop-and-go waves because of a downstream incident involving multiple lane closures. Excluding these outliers, the average daily \ce{CO2} reduction potential on a typical weekday falls between 10.5\% and 17.5\%. These observations suggest that the severity and frequency of stop-and-go waves on a given day significantly influence the emission reduction potential of wave-smoothing strategies, which can be a topic for future further investigation.

\section{Discussion}
\label{sec:discussion}

\subsection{Trade-off between gap size and emission reduction}
\begin{figure}[H]
    \centering
    \includegraphics[width=0.75\linewidth]{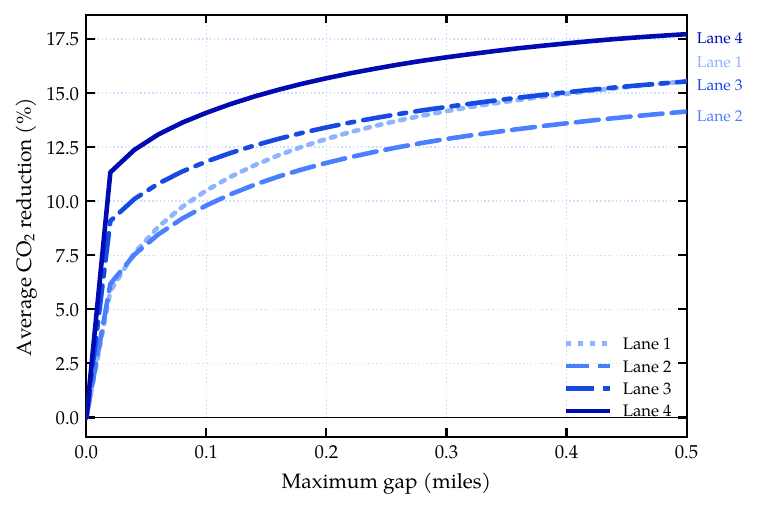}
    \caption{\textbf{Trade-off between maximum gap size and \ce{CO2} emission reduction by lane.} The figure illustrates how the maximum-gap constraint affects \ce{CO2} emission reductions across lanes, highlighting the balance between traffic-flow smoothness and emission benefits.}
    \label{fig:trade-off-gap}
\end{figure}

We examine the relationship between average emission reduction potential and maximum gap size across all days. We first consider the \ce{CO2} reduction potential because, in the MOVES model, a reduction in \ce{CO2} emissions is equivalent to fuel savings. The results suggest that minor adjustments to driving behavior during congestion can yield substantial improvements: an allowable gap of only 0.02 miles (approximately 32 meters) can yield a 6\% to 11\% reduction in \ce{CO2} emissions across lanes (as shown in \figurename~\ref{fig:trade-off-gap}). Similar patterns are observed for the other pollutants (CO, HC, and \ce{NOx}). The detailed trade-off curves for all four pollutants are presented in ~\ref{sec:appendix-sensitivity} (\figurename~\ref{fig:tradeoff-gap-all}).

\subsection{Emission reduction potential across lanes}
\label{sec:lane-comparison}
\begin{figure}[H]
    \centering
    \includegraphics[width=0.80\linewidth]{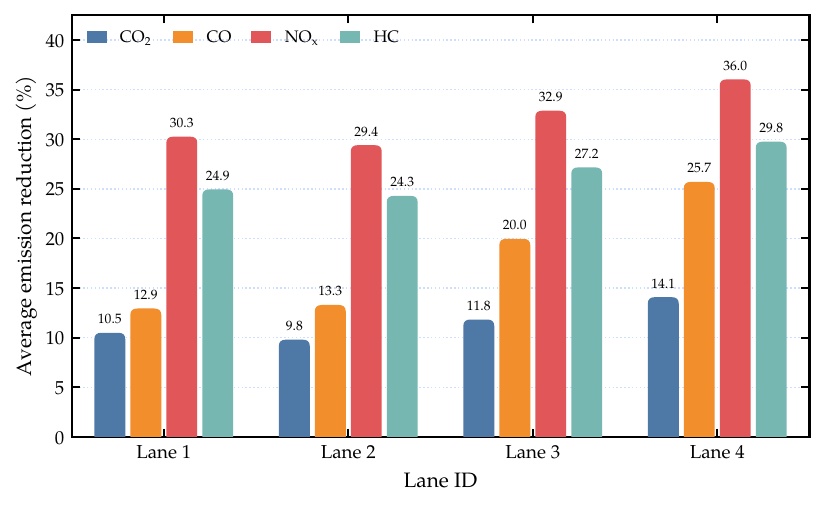}
    \caption{\textbf{Lane-by-lane emission reduction comparison under a 0.1-mile maximum-gap constraint.} Average emission reduction percentages for \ce{CO2}, CO, \ce{NOx}, and HC in each lane show that the rightmost lanes (Lanes~3 and~4) generally achieve greater reductions than the leftmost lanes (Lanes~1 and~2).}
    \label{fig:lane-comparison}
\end{figure}

As lane-level congestion patterns vary~\cite{ji2026scalable}, we next compare the emission-reduction potential across lanes under a moderate maximum-gap constraint of 0.1~mile. \figurename~\ref{fig:lane-comparison} shows that the rightmost lanes (Lanes~3 and~4) consistently achieve higher emission reductions than the leftmost lanes (Lanes~1 and~2), by roughly 16\% for \ce{NOx} and HC, 28\% for \ce{CO2}, and 74\% for CO. Consistent with our earlier findings on lane-level wave dynamics~\cite{ji2026scalable}, the rightmost lanes typically exhibit less severe waves and smaller speed-oscillation amplitudes than the leftmost lanes.

\subsection{Emission reduction potential across vehicle classes}
\begin{figure}[H]
    \centering
    \includegraphics[width=0.75\linewidth]{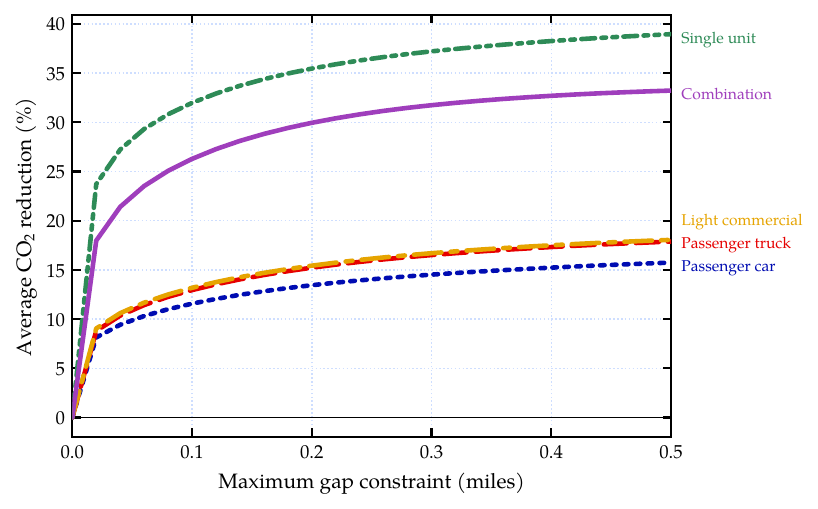}
    \caption{\textbf{Trade-off between maximum gap size and \ce{CO2} emission reduction by vehicle class.} Average \ce{CO2} reduction as a function of the maximum-gap constraint for the five vehicle classes specified in \tablename~\ref{tab:classspec}, averaged over all benchmarked days and lanes. The two heavy-duty truck classes (single-unit and combination short-haul) show roughly two to three times the reduction potential of the three light-duty classes, while all classes follow the same saturating trend.}
    \label{fig:vehicle-class}
\end{figure}

The results reported so far assume a passenger car. We repeat the same analysis for the five vehicle classes listed in \tablename~\ref{tab:classspec} and report the resulting \ce{CO2} trade-off curves in \figurename~\ref{fig:vehicle-class}.

All five classes follow the same trade-off pattern reported earlier: the reduction rises steeply within the first 0.02~mile of allowable gap and then flattens, with roughly half to 60\% of the benefit attainable at 0.5~mile already realized at 0.02~mile, regardless of the vehicle considered. The magnitude, however, differs considerably across classes. The three light-duty gasoline classes behave almost identically, reaching about 11\% (passenger car) to 13\% (passenger truck and light commercial) under the 0.1-mile constraint used throughout this study. The two heavy-duty diesel classes are far higher, at about 26\% for combination short-haul trucks and about 32\% for single-unit short-haul trucks, and both remain roughly two to three times above the light-duty classes across the entire range of gap sizes. In other words, the heavier the vehicle, the more of its emissions are attributable to stop-and-go waves and the more it stands to gain from smoothing.

This reinforces the lane-level finding in \sectionname~\ref{sec:lane-comparison}: the rightmost lanes both carry a larger share of freight trucks and offer greater reduction potential. 
This result is particularly relevant because Lanes~3 and~4 carry a larger proportion of commercial freight trucks. Truck operators are plausible early adopters of wave-smoothing policies, and commercial trucks are practical candidates for deployment and field testing as they usually have well-trained drivers and more advanced driver-assistance systems.
Trucks are therefore doubly attractive targets for wave-smoothing strategies: they travel in the lanes where the potential is highest, and they benefit the most per vehicle.

\subsection{Implications for traffic management and control strategies}
The analysis in this study provides evidence-based implications for traffic management and control strategies in the presence of stop-and-go waves, as summarized below.

\textit{Moderate increases in gap size yield meaningful emission reductions}. As shown in \figurename~\ref{fig:trade-off-gap}, even a maximum gap of 0.02 miles yields meaningful emission reductions when stop-and-go waves are not fully mitigated. This finding suggests that traffic management strategies that promote slightly larger headway are broadly beneficial.

\begin{figure}[H]
    \centering
    \includegraphics[width=0.65\linewidth]{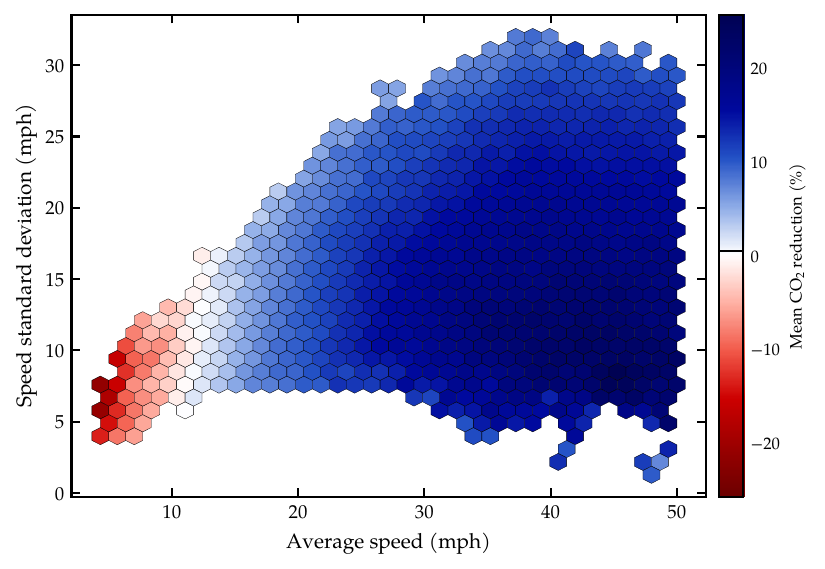}
    \caption{\textbf{\ce{CO2} emission reduction potential for different trajectory mean speeds and speed standard deviations.} Each hexagonal bin represents the average \ce{CO2} reduction potential for trajectories within specific ranges of mean speed and speed standard deviation. The color intensity indicates the magnitude of the reduction potential.}
    \label{fig:hexbin}
\end{figure}

\textit{Smoothing is more effective when waves are addressed early}. As the intensity of traffic waves increases, particularly in lane-closure bottleneck scenarios, the emission reduction potential decreases significantly (\figurename~\ref{fig:day-to-day}). As shown in \figurename~\ref{fig:hexbin}, the reduction potential approaches 25\% for trajectories with a high mean speed and a moderate speed standard deviation. Somewhat counterintuitively, the reduction potential becomes marginal, and can even turn negative, for trajectories with a low mean speed, because the deep, sustained slowdowns those trajectories experience cannot be effectively smoothed while maintaining a reasonable headway. This finding suggests that traffic management strategies should prioritize early intervention to smooth traffic waves before they intensify, thereby maximizing emission reduction benefits.

\textit{The benchmark translates naturally into a real-time control formulation}. Because this study is an offline evaluation, the smoothed trajectory is planned directly over the entire observation horizon. For real-time deployment~\cite{coifman2026using, yang2014control, jang2025reinforcement}, the same optimization pipeline can be recast as a model predictive control problem, in which the control action is the suggested speed over a finite planning horizon, a prediction of the vehicle's future trajectory serves as the input, and the solution is updated in a rolling-horizon manner. Under this formulation, the maximum-gap constraint is interpreted as the additional gap that the controlled vehicle is permitted to open relative to the gap it would otherwise maintain, which is directly implementable either as a spacing bound in an automated controller or as speed guidance provided to a driver. Such control would still require supervision by trained drivers to ensure safety. 

\subsection{Generalizability}
\begin{figure}[H]
    \centering
    \includegraphics[width=0.8\linewidth]{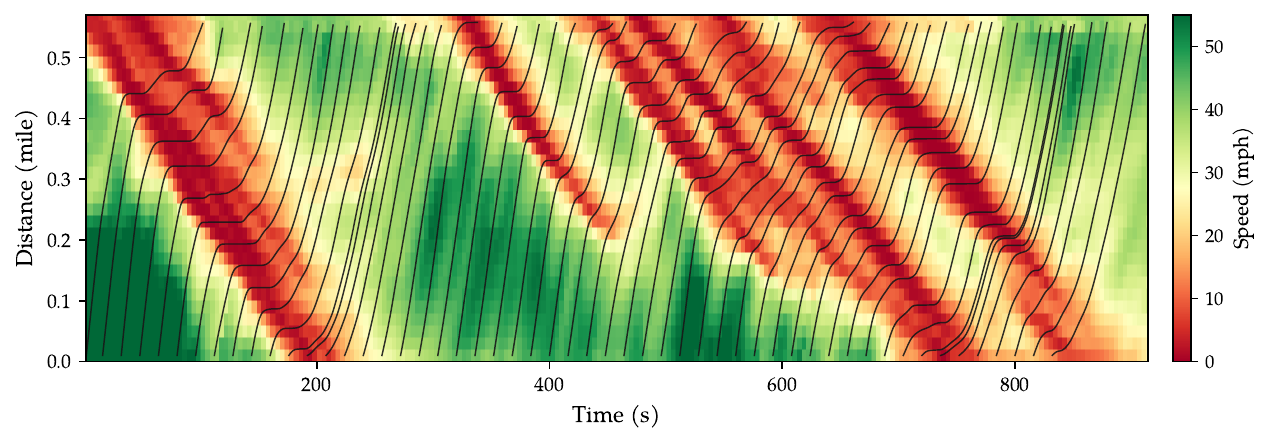}
    \caption{\textbf{Stop-and-go waves in the MiTra A50 dataset.} Time--space diagram of the A50 recording (raw data in its flight 4, lane 4), with the speed field color-coded from red (congestion) to green (free flow) and a subset of virtual trajectories overlaid in black. Several stop-and-go waves propagate upstream within the 0.54-mile observation window, qualitatively similar to the wave dynamics observed on the I-24 MOTION testbed.}
    \label{fig:mitra}
\end{figure}

To demonstrate how our method can be transferred to other traffic environments, we applied it to MiTra~\cite{chaudhari2025mitra}, a recently released trajectory dataset collected in Italy that exhibits similar stop-and-go traffic dynamics, as shown in \figurename~\ref{fig:mitra}.

\begin{figure}[H]
    \centering
    \includegraphics[width=1\linewidth]{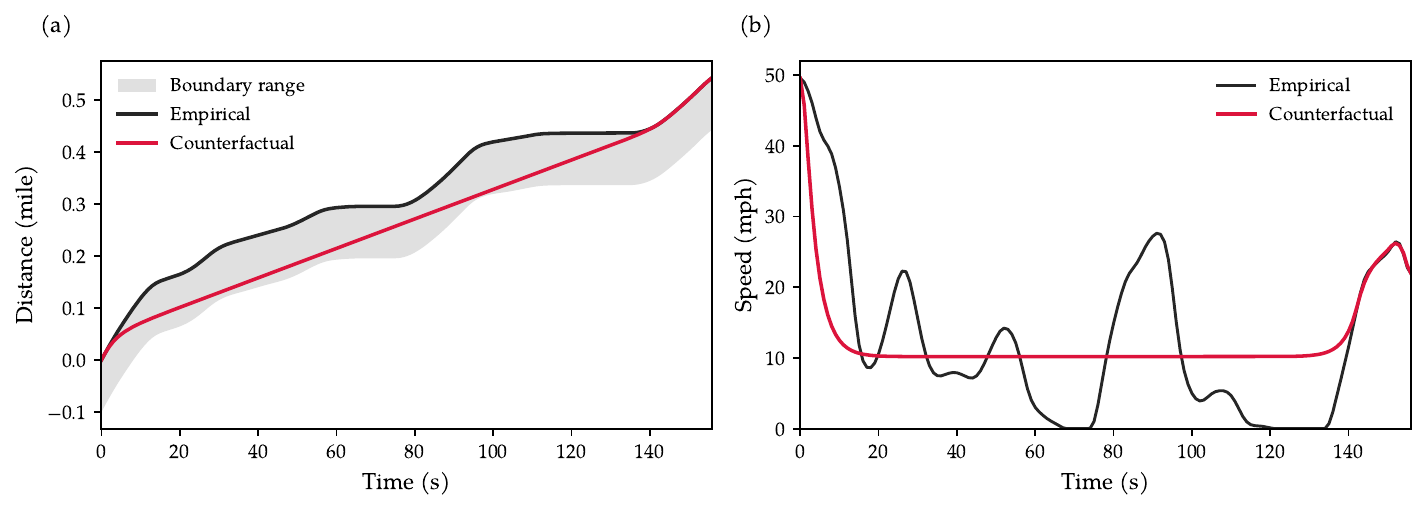}
    \caption{\textbf{Counterfactual wave-smoothing benchmark applied to a MiTra A50 trajectory.} (a) Time--space diagram of an empirical A50 trajectory and its benchmark counterpart under a 0.1-mile maximum-gap constraint, with the admissible boundary range shaded. (b) The corresponding speed profiles.}
    \label{fig:a50mitra}
\end{figure}

We repeat the same pipeline on the MiTra A50 data~\cite{chaudhari2025mitra}: for each empirical trajectory we reconstruct a counterfactual benchmark trajectory by solving the same quadratic program with fixed boundary conditions and a maximum-gap constraint, and we then estimate the emissions of both trajectories with MOVES. \figurename~\ref{fig:a50mitra} shows one empirical A50 trajectory together with its benchmark counterpart under a 0.1-mile maximum-gap constraint. The benchmark behaves exactly as it does on I-24 MOTION: it mitigates the stop-and-go oscillations while remaining inside the admissible boundary range and preserving the entry and exit conditions. This confirms that the framework transfers to a different site without reformulation.

\begin{figure}[H]
    \centering
    \includegraphics[width=\linewidth]{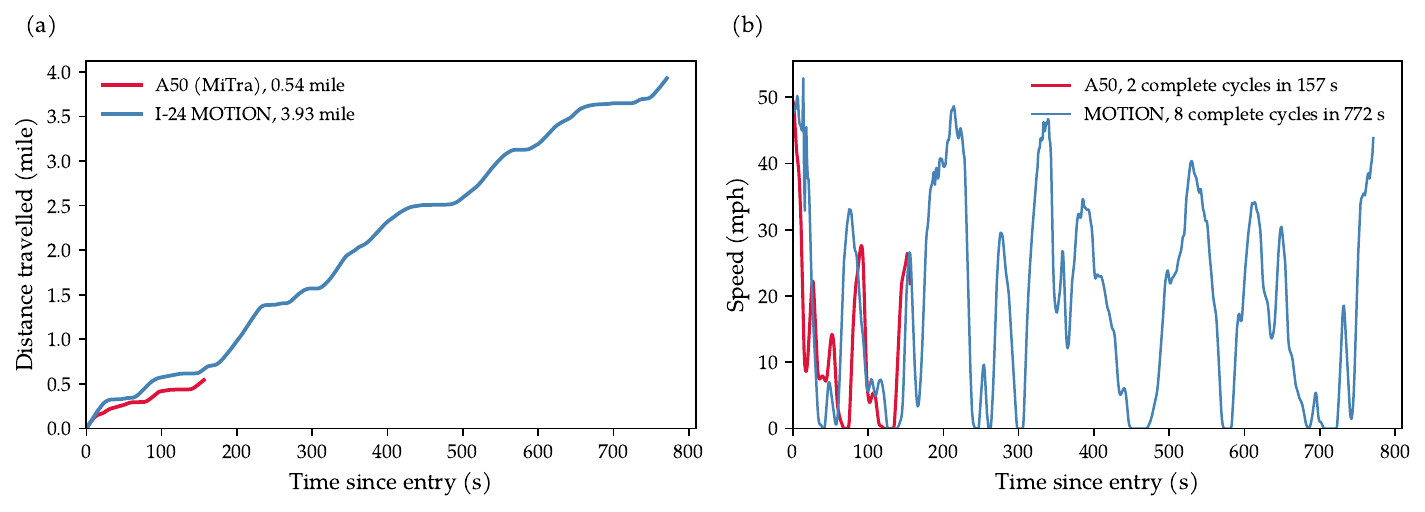}
    \caption{\textbf{Comparison of the observed driving cycles in MiTra and I-24 MOTION.} (a) Distance travelled since entering the observation section for an exemplar trajectory from each dataset. (b) The corresponding speed profiles. The A50 vehicle covers 0.54 mile in 157~s and experiences two complete stop-and-go cycles, while the I-24 MOTION vehicle covers 3.93 miles in 772~s and experiences eight.}
    \label{fig:mitramotion}
\end{figure}

The two sites differ substantially in the length of the driving cycle they observe, which is the main reason the two analyses are not directly comparable in magnitude. \figurename~\ref{fig:mitramotion} contrasts an exemplar trajectory from each dataset. Because the A50 section is only 0.54 mile long, a vehicle traverses it in about 157~s and experiences roughly two complete stop-and-go cycles, whereas a virtual trajectory observed over the 3.93-mile I-24 MOTION section is tracked for about 772~s and experiences eight. Each A50 trajectory therefore captures only a few wave cycles, and a comparatively large share of it is taken up by the entry and exit transients that the fixed boundary conditions preserve and that smoothing cannot act on.

\begin{figure}[H]
    \centering
    \includegraphics[width=0.75\linewidth]{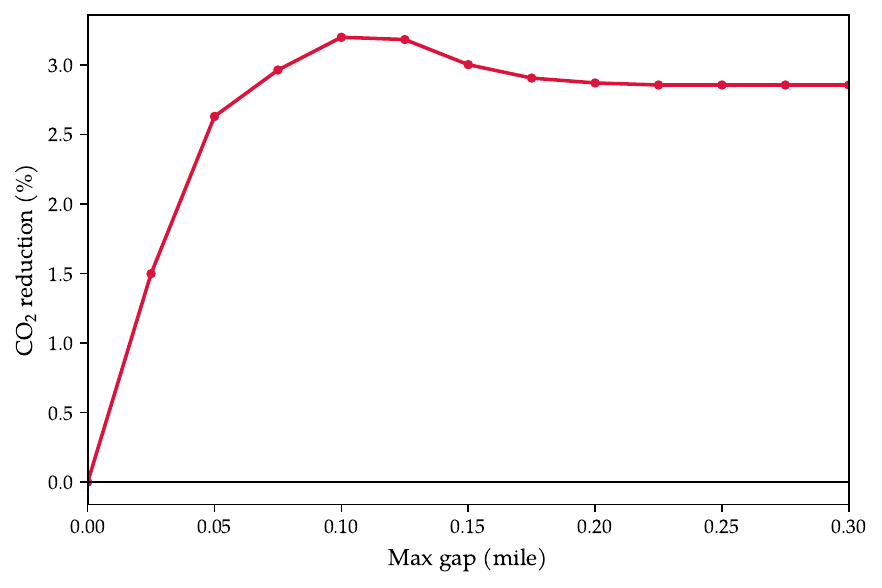}
    \caption{\textbf{\ce{CO2} reduction potential versus the maximum allowable gap on the MiTra A50 data.} The curve is the average over 211 A50 trajectories.}
    \label{fig:mitragap}
\end{figure}

The aggregated results reflect this difference. \figurename~\ref{fig:mitragap} reports the \ce{CO2} reduction potential for passenger cars of the benchmark as a function of the maximum allowable gap across the 211 A50 virtual trajectories we tested in \figurename~\ref{fig:a50mitra}. The shape of the trade-off curve is consistent with what we report for I-24 MOTION: most of the attainable benefit is realized at small gaps, and further relaxation of the constraint adds little. The magnitude, however, is considerably smaller: about 3.2\% at a 0.1-mile maximum gap, settling near 2.9\% beyond it, compared with 9.80\% to 14.08\% across lanes on I-24 MOTION. This is expected given the short observation window: individual trajectories that happen to contain a full wave benefit substantially, while trajectories dominated by boundary transients show little or slightly negative benefit. We report these values only to show that the pipeline runs end to end on an independent dataset, not as a cross-site quantitative comparison.

Note that because MiTra covers only a few hours of traffic over 0.54 miles (roughly 1.5 lane-mile-hours), compared with roughly 700 lane-mile-hours analyzed in our primary study, it is not sufficiently large to support a comprehensive cross-site quantitative comparison.

Regarding the factors that may affect the evaluation results across sites, we identify two principal ones. The first is the wave pattern itself. Wave amplitude, duration, and frequency jointly determine the proportion of a trip that a vehicle spends in the deceleration and re-acceleration phases of a wave, and therefore how much of its emissions is attributable to the waves and can be removed by smoothing in the first place. This is precisely the mechanism behind the day-to-day variation we report for I-24 MOTION, where the daily average \ce{CO2} reduction ranges from about 2.5\% to 17.5\% depending on wave severity, and behind the lane-to-lane differences we observe. A site with different wave characteristics should therefore be expected to exhibit a different reduction potential even under an identical smoothing strategy, which is why we report the potential as a distribution and as a function of the maximum allowable gap rather than as a single transferable number. The second is the local emission rate. We use MOVES throughout, including for the A50 analysis, because it is the standard tool for project-level analysis in the United States and, to the best of our knowledge, no comparable openly available emission model with vehicle-specific-power-based operating-mode rates exists for the European fleet. To best evaluate other site emission reduction potential of stop-and-go wave smoothing, a local emission rate would capture the effect more precisely.

\subsection{Limitations and future work}
The empirical trajectories analyzed in this paper are virtual trajectories reconstructed from high-resolution mean speed fields; therefore, individual driving signals are averaged and smoothed. Such trajectories have been shown to be well suited to emission estimation~\cite{tsanakas2022generating}, and they make an analysis of hundreds of thousands of trajectories possible. Pairing the benchmark with individually tracked trajectories would additionally reveal how driver-to-driver variability shapes the reduction potential.

The benchmark is computed for one trajectory at a time, which isolates the potential attributable to a single smoothing vehicle and therefore provides a conservative reference point. Collective effects are a natural extension: when a platoon of vehicles adopts the same policy, the achievable reduction may be considerably larger~\cite{coifman2026using}. In addition, the benchmark acts mainly on longitudinal dynamics, consistent with the mechanism of stop-and-go waves; the reported values should therefore be interpreted as the potential of longitudinal wave smoothing alone. A recent study shows that stop-and-go waves can also propagate laterally across lanes~\cite{li6172812comprehensive}. Extending the benchmark to include both longitudinal and lateral dynamics offers a promising avenue for future research: reducing unnecessary lane changes would both eliminate the acceleration transients of the lane-changing vehicle and dampen lateral wave propagation, potentially producing greater emission savings than longitudinal smoothing alone.

On the emission estimation side, particulate matter (PM) would be a particularly valuable addition because it is closely associated with braking-related emissions, whereas the pollutants considered here respond less strongly to braking. PM will also remain a concern as the vehicle fleet electrifies, making it a natural target for wave-smoothing strategies. The tool developed in this study is designed to support these extensions and provides a foundation for evaluating a broader and more complex set of scenarios.

\section{Conclusion}
\label{sec:conclusion}
In summary, this is the first study to analyze the emission reduction potential of large-scale freeway traffic smoothing using a high-resolution empirical dataset. Our analysis is three orders of magnitude larger than those in prior studies and covers 40 days and more than 500,000 empirical vehicle trajectories collected by the I-24 MOTION testbed, a large-scale freeway traffic observation system. \textbf{We note that these trajectories are virtual trajectories reconstructed from high-resolution speed fields, rather than directly observed raw vehicle trajectories.} We develop a convex-optimization-based trajectory-smoothing approach to construct smooth, feasible benchmark trajectories from empirical trajectories and estimate their emission reduction potential using the U.S. EPA MOVES model.
Unlike prior studies, we impose a maximum-gap constraint to ensure that the benchmark trajectories are feasible under real-world traffic conditions.
The results show that smoothing stop-and-go waves can yield meaningful emission reductions across lanes under a moderate maximum-gap constraint of 0.1 miles: average reductions of 9.80\% to 14.08\% for \ce{CO2}, 12.94\% to 25.70\% for CO, 24.29\% to 29.76\% for HC, and 29.40\% to 36.03\% for \ce{NOx}. These findings highlight the environmental benefits of traffic-wave mitigation and provide valuable insights for traffic management and control strategies.

Building on this work, we envision field-testing additional wave-smoothing strategies in real-world traffic. Combining observations of empirical traffic-wave dynamics from the I-24 MOTION testbed with the emission estimation tools developed in this study will allow us to quantify the real-world emission reduction potential of wave smoothing, particularly the system-level benefits that may emerge when many vehicles adopt these strategies and collectively smooth traffic flow. The full potential of wave-smoothing strategies will be unlocked through further real-world experiments and deployments --- a prospect we are excited to pursue.

\section*{Acknowledgments}
We appreciate the editor and the anonymous reviewers for their constructive comments and suggestions, which have helped improve the quality of this paper. The motivation and ideas presented here grew out of numerous field experiments conducted in Fall 2025. We thank Dr. Matt Bunting at Vanderbilt University for his support and involvement in those experiments. We also appreciate Professor Hiba Baroud and Professor Ahmad Taha at Vanderbilt University for their feedback on this work.
This work was supported by the Tennessee Department of Transportation (TDOT) under Grant No. OTH2023-01F-01, the TDOT University Technical Assistance Program under Grant No. 402526, the National Science Foundation (NSF) under Grant Nos. 2434400 (Work, Sprinkle) and 2111688 (Sprinkle), and the Vanderbilt University Lacy-Fischer Interdisciplinary Research Grant (Ji) and Dissertation Enhancement Grant (Ji). We also acknowledge the NVIDIA Academic Grant Program (Work, Ji) for providing computational resources. The views expressed herein are those of the authors and do not necessarily reflect those of the Tennessee Department of Transportation or the United States Government.

\pagebreak
\appendix
\section{Trade-off analysis of gap size and emission reduction}
\label{sec:appendix-sensitivity}
\begin{figure}[H]
    \centering
    \includegraphics[width=0.75\linewidth]{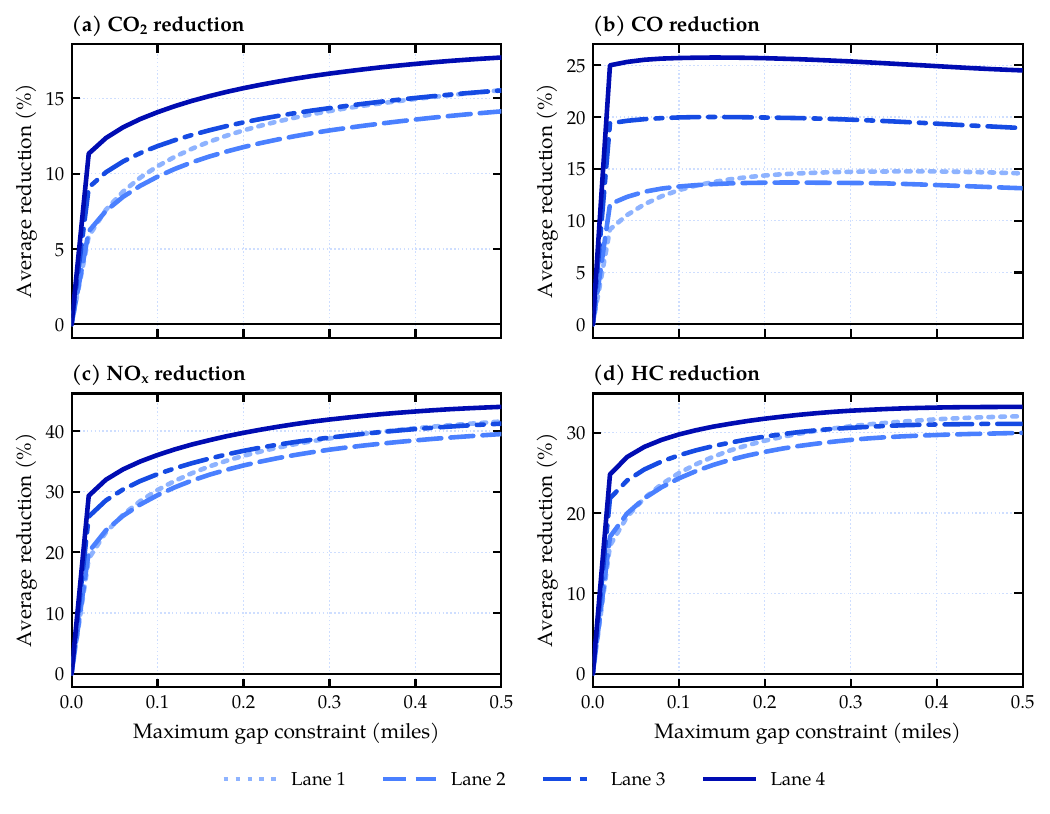}
    \caption{Trade-off between maximum gap size and emission reduction potential across lanes for \ce{CO2}, \ce{CO}, \ce{NOx}, and \ce{HC}.}
    \label{fig:tradeoff-gap-all}
\end{figure}

\begin{figure}[H]
        \centering
        \includegraphics[width=0.75\linewidth]{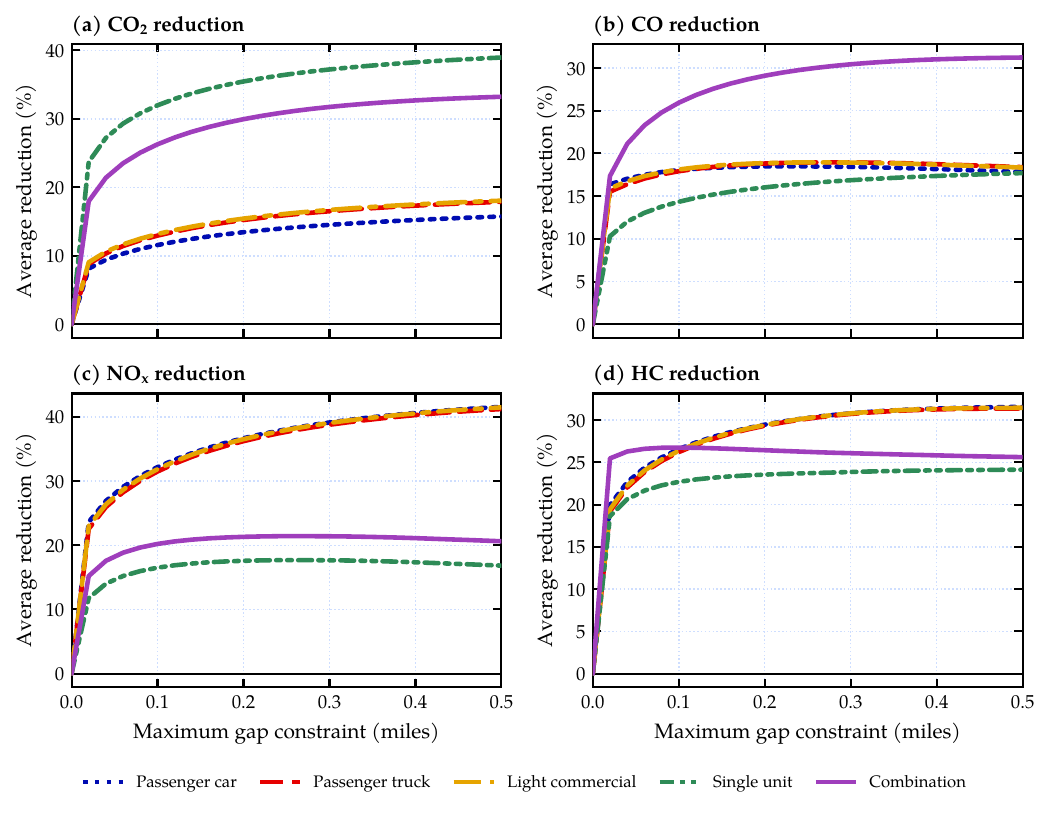}
        \caption{Trade-off between maximum gap size and emission reduction potential across vehicle classes for \ce{CO2}, \ce{CO}, \ce{NOx}, and \ce{HC}.}
        \label{fig:trade-off-vehicle-class}
\end{figure}

\figurename~\ref{fig:tradeoff-gap-all} presents the trade-off curves between maximum gap size and emission reduction potential across lanes for all four pollutants (\ce{CO2}, CO, \ce{NOx}, and HC). \figurename~\ref{fig:trade-off-vehicle-class} presents the corresponding curves across the five vehicle classes of \tablename~\ref{tab:classspec}, extending the \ce{CO2} results of \figurename~\ref{fig:vehicle-class} to the other three pollutants. \figurename~\ref{fig:trade-off-vehicle-class} presents the corresponding curves across the five vehicle classes of \tablename~\ref{tab:classspec}, extending the \ce{CO2} results of \figurename~\ref{fig:vehicle-class} to the other three pollutants.

\section{Why emissions respond differently to gap opening}
\label{sec:appendix-mechanism}
\begin{figure}[H]
    \centering
    \includegraphics[width=0.80\linewidth]{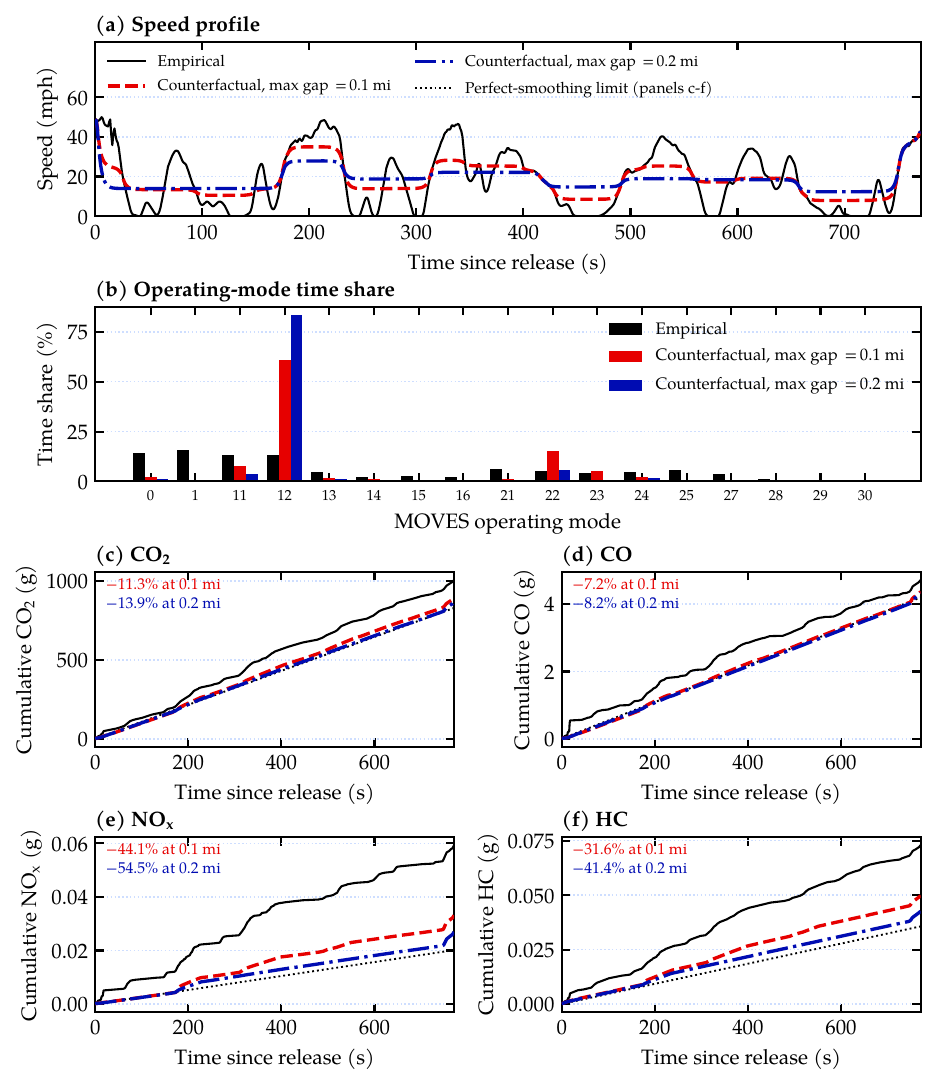}
    \caption{Effect of trajectory smoothing on vehicle emissions across two maximum-gap scenarios. (a) Speed profile over time since release, comparing the empirical trajectory against counterfactual smoothed trajectories at maximum gaps of 0.1 mi and 0.2 mi, alongside the perfect-smoothing limit used in panels (c to f). (b) Time share across MOVES operating modes, which is identical across all four pollutants shown in (c to f). (c to f) Cumulative emissions over time for \ce{CO2}, \ce{CO}, \ce{NOx}, and \ce{HC}, respectively, with percent reductions relative to the empirical trajectory shown for each counterfactual scenario: \ce{CO2} decreases by 11.3\% (0.1 mi) and 13.9\% (0.2 mi); \ce{CO} by 7.2\% and 8.2\%; \ce{NOx} by 44.1\% and 54.5\%; and \ce{HC} by 31.6\% and 41.4\%.}
    \label{fig:understanding}
\end{figure}

\figurename~\ref{fig:understanding} illustrates why the four pollutants respond so differently to the same relaxation of the maximum-gap constraint. The difference arises from the operating-mode distribution shift shown in \figurename~\ref{fig:understanding}(b), combined with how emission rates vary across modes in the MOVES database. Opening the gap from 0.1 to 0.2 miles shifts more time from high-emitting modes (21 to 30) to lower-emitting cruise/coast modes (11 to 16). The MOVES emission rate tables show that \ce{CO2} and CO rates are relatively uniform across these modes, so the shift produces only a modest change in their cumulative emissions, 2.6 and 1.0 percentage points respectively. \ce{NOx} and \ce{HC} rates, in contrast, vary much more sharply across modes, so the same distributional shift yields substantially larger additional reductions, 10.4 and 9.8 percentage points. The same asymmetry explains the ordering at a fixed gap: at 0.1 mile the benchmark already removes 44.1\% of \ce{NOx} and 31.6\% of HC but only 11.3\% of \ce{CO2} and 7.2\% of CO.

\bibliographystyle{ieeetr}
\bibliography{refs.bib}
\end{document}